\documentclass{optica-article}

\journal{opticajournal} % for journals or Optica Open

\articletype{Research Article}

\usepackage{lineno}
\renewcommand{\normalsize}{\fontsize{12}{14}\selectfont}
\usepackage[version=4]{mhchem}
\usepackage{caption}
\usepackage{subcaption}
\usepackage{float}
\usepackage{cleveref}
\usepackage{xspace}
\usepackage{amsmath}
\usepackage{setspace}
\newcommand{\ie}{\textit{i}.\textit{e}., }

\newcommand{\via}{\textit{via}\;}
\newcommand{\etal}{\textit{et al}., }
\newcommand{\hmfe}{Half Maxwell Fish Eye\xspace}
\newcommand{\lbdao}{\lambda_0}
\newcommand{\mum}{\,$\mu$m\xspace}
\newcommand{\nm}{\,nm\xspace}
\newcommand{\neff}{n_{eff}}
\newcommand{\neffphc}{n_{eff}^{PhC}}
\newcommand{\neffphcmode}{n_{eff}^{PhC-mode}}

\newcommand{\Sio}{\ce{SiO2}\xspace}
\newcommand{\nsio}{n_{\Sio}\xspace}
\newcommand{\Si}{\ce{Si}\xspace}% bulk Si
\newcommand{\instantE}{\lvert\,E\,\rvert}
\newcommand{\intens}{{\lvert\,E\,\rvert}^2}
\newcommand{\intensdis}{field intensity distribution $\intens$\xspace}
\newcommand{\Intensdis}{Field intensity distribution $\intens$\xspace}
\newcommand{\eintensdis}{electric field intensity distribution\xspace}

\newcommand{\teo}{$\mathbb{TE}_0$\xspace}% TE0 mode
\newcommand{\teu}{$\mathbb{TE}_1$\xspace} % TE1 mode
\newcommand\xyplane{$x\,y$\protect\nobreakdash-plane\xspace}
\newcommand\yzplane{$y\,z$\protect\nobreakdash-plane\xspace}
\newcommand{\seef}[1] {(see~Fig.\,\ref{#1})} % {(see~figure\:\ref{#1})} --- faon scientific  reports
\newcommand{\seefd}[2] {(see~Fig.\,\ref{#1} and \ref{#2})}
\newcommand{\Fig}[1] {Fig.\,\ref{#1}} % {(see~figure\:\ref{#1})} --- faon scientific  reports
\newcommand{\seeeq}[1]{(see~eq.\eqref{#1})}

\newcommand{\seesec}[1]{(see~section\:\ref{#1})}
\usepackage{hyperref}
\usepackage{xr}
\usepackage{xr-hyper}
\usepackage{hyperref}
\date{\today}

\begin{document}
%\date{\today}
\title{Broadband on-chip Half Maxwell Fisheye}

%\author{Author One,\authormark{1} Author Two,\authormark{2,*} and Author Three\authormark{2,3}}

\author{Xin ZHENG\authormark{1}, Quan YUE\authormark{1}, Jean-Ren{\'e} COUDEVYLLE\authormark{1}, Aziz BENAMROUCHE\authormark{2}, S{\'e}gol{\`e}ne CALLARD\authormark{2}, Anatole LUPU\authormark{1}, \'Eric AKMANSOY\authormark{1,*}
}

%\address{\authormark{1}Peer Review, Publications Department, Optica Publishing Group, 2010 Massachusetts Avenue NW, Washington, DC 20036, USA\\
%\authormark{2}Publications Department, Optica Publishing Group, 2010 Massachusetts Avenue NW, Washington, DC 20036, USA\\
%\authormark{3}Currently with the Department of Electronic Journals, Optica Publishing Group, 2010 Massachusetts Avenue NW, Washington, DC 20036, USA}

\address{\authormark{1} Centre de Nanosciences et Nanotechnologies, CNRS, UMR 9001,  Universit\'e Paris-Sud, Universit{\'e} Paris-Saclay, C2N-Orsay, 91120, Palaiseau, France\\
\authormark{2}  \'Ecole Centrale de Lyon, INSA Lyon, CNRS, Universit\'e Lyon 1, CPE Lyon, INL, UMR5270, 69130 \'Ecully, France
}

\email{\authormark{*}eric.akmansoy@universite-paris-saclay.fr} %% email address is required; see note below about the corresponding author designation

% use {asbstract*} to suppress the copyright line. Copyright information will be added in production

\begin{abstract*} 
In this article, we report on the design and the experimental evidence of a Half Maxwell Fish Eye (HMFE), for Silicon Photonics and working at telecommunication wavelength. %The HMFE is derived from the Maxwell Fish Eye ???
It is designed by implementing a Graded Photonic Crystal operating in the non-resonant metamaterial regime. The results of 3D Finite-Difference Time-Domain simulations (FDTD) show an excellent broadband focusing capacity. It has been further fabricated \via the Silicon On Insulator (SOI) platform for its compatibility with CMOS technology. Experimentally, its performances are firstly  investigated by the means of a fan-shaped set output waveguides. Next, Scanning Near-Field Optical Microscopy (SNOM) characterisation confirms the wavefront curving inside the HMFE lens. Quantitative analysis of the SNOM results demonstrates its excellent focusing performances: the Full Width Half Maximum (FWHM) is $0.466\lbdao$ at $\lbdao$=1550\nm, while the thickness of the lens is 3.18$\lbdao$.
\end{abstract*}
%  validated the excellent focalising capacity of this lens 

%%%%%%%%%%%%%%%%%%%%%%%%%%  body  %%%%%%%%%%%%%%%%%%%%%%%%%%
\section{Introduction}
\begin{figure}[htbp]
    \centering
    \begin{subfigure}[c]{0.85\textwidth} % "0.45" donne ici la largeur de l'image
        \centering \includegraphics[width=0.5\textwidth]{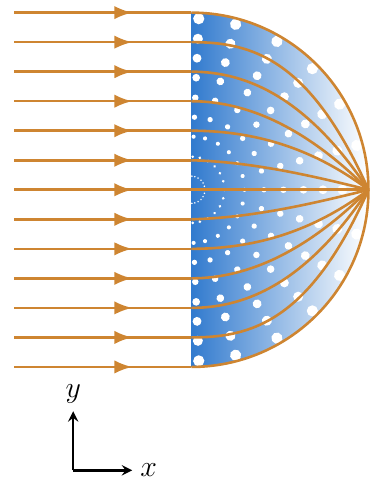}
        \caption{}\label{fig:principle}
    \end{subfigure}
\hfill    ~ % ce symbole ajoute un espacement horizontal entre les premires deux images

 % la ligne blanche correspond au retour ˆ la ligne aprs le deuxime image
    \begin{subfigure}[c]{0.85\textwidth}
        \centering \includegraphics[width=\textwidth]{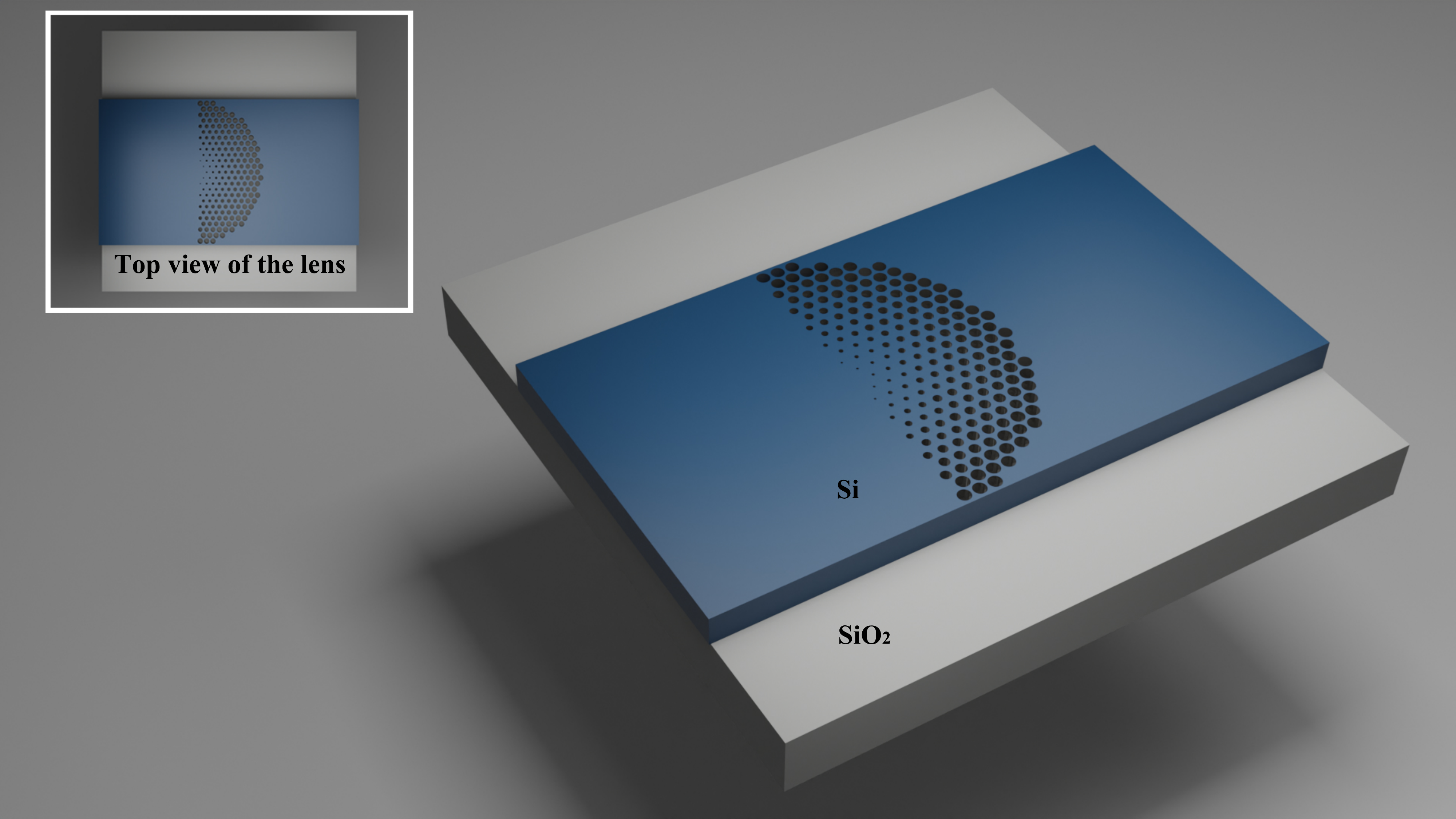}
        \caption{}\label{fig:fig1b}
    \end{subfigure}
       \caption{(a) Sketch of the 2D Half Maxwell Fish Eye (HMFE) implemented \via a GPC, whose air filling factor increases from the center towards the rim. The incident plane wave is focused onto its output surface. (b) Schematic diagram of the designed HMFE on a SOI chip for integrated optics. The HMFE is fed by a strip waveguide. Inset: top view of the chip  (\xyplane).}\label{fig:schematic}
  \end{figure}

 The interest in miniaturising different photonic components continues to grow and has attracted the attention of many researchers.\cite{xiang2021perspective,engheta2007circuits,thomson2016roadmap} The miniaturised devices at telecommunication wavelengths help to increase the level of integration, improving the performance of the Photonic Integrated Circuits (PIC). GRadient INdex (GRIN) optics is the mean for photonic devices miniaturisation, because it allows the design of any index profile.\cite{luque2019ultracompact, zhang2020ultra} It is remarkable that GRIN lenses may be free of aberrations.\cite{gomez2012gradient, bociort1994imaging, book_bornwolf} The Maxwell FishEye (MFE) is one of the most famous GRIN lenses in history.\cite{maxwell1854solutions} It is an absolute optical instrument because it forms perfect images, from the perspective of geometric optics. It is a spherically symmetric lens and stigmatic: all the rays emerging from one point inside it converge to a unique point.\cite{njp13_tyc} The path of the internal optical rays are along arcs of circles.\cite{book_bornwolf} Finding out if the MFE is a super-resolution lens, i.e., it can resolve details much smaller that the wavelength, has been a matter of controversy (see \cite{n7375_tyc_zhang} and references therein). 
The Half Maxwell Fish-Eye (HMFE), as its name suggests, is the half of the Maxwell Fish-Eye, a sketch of which is shown in \Fig{fig:principle}. It retains the MFE optical properties. %, which is an absolute instrument of variable refractive index. \cite{book_bornwolf} 
It is a hemisphere, and it is a useful device for photonic applications because it collimates any point source located on its spherical rim into a directive plane wave whose aperture is that of the lens. Reciprocally, it focuses a plane wave incident on its plane surface onto its output spherical surface, which is what we are interested in here. For designing PICs for integrated photonics, a 2D HMFE is implemented, as a miniaturized GRIN lense.\cite{luque2019ultracompact, fan2017integrated, zhang2020ultra} %will be fabricated in the \Si layer of a . 

However, the fabrication of these components in the optical domain remains challenging.\cite{ao19_moore} Thanks to the recent advancement of nanofabrication processes, the fabrication of the MFE and HMFE has been carried out at microwave, terahertz, and even telecommunication wavelengths. Lei \etal fabricated a MFE using all-dielectric polymer/ceramic composite materials for the microwave field, whose GRIN profile is achieved by varying its composition.\cite{lei2017generalized} Headland \etal varied the air hole size of the Photonic Crystal (PhCs) to get the GRIN profile and fabricated the HMFE at the terahertz range.\cite{headland2020half,headland2021dielectric} Bitton \etal succeeded in fabricating a MFE at telecommunication wavelengths and the GRIN profile is obtained by varying the thickness of the guiding layer.\cite{bitton2018two} Furthermore, Shen \etal have demonstrated a HMFE fiber-to-chip coupleur. \cite{nanoph11_shen}

In this article, we report on the design, the fabrication and the characterisation of a 2D HMFE fabricated on a Silicon On Insulator (SOI) chip for telecommunication wavelengths ($\lbdao = 1.55$\mum). 
The HMFE is implemented \via a Graded Photonic Crystal (GPC), which is realized by gradually modifying one parameter of the unit cell of PhCs.\cite{centeno2005graded} It operates in the non-resonant metamaterial regime \ie\;the lattice period satisfy $a < \lambda/2\neff$. The PhC then behaves as a homogeneous and isotropic medium, since the Iso-Frequency Curves (IFC) --- that represent the relation of dispersion at a given frequency ---,\cite{oc285_gaufillet, om47_gaufillet}  are almost circular; it is even almost dispersionless.\cite{staude2017metamaterial, cheben2018subwavelength}
The 2D radially graded index $n(r)$ of the HMFE is realized \via the gradient of the effective index $\neff$ of a GPC, which is achieved by varying the air filling factor $\eta$ over the unit cell of PhCs. \cite{pj10_gaufillet} 
Once the HFME designed, 3D Finite-Difference Time-Domain (FDTD) simulations are carried out to verify its optical properties, and they show that it focuses an incident plane wave over a broad band of wavelengths. Next, the HMFE is fabricated by the means of CMOS technology on a SOI platform in the C2N clean room, and then characterised in two steps. The first one involving a fan-shaped set of output waveguides confirms the focusing property of the fabricated device. The second one using Scanning Near-Field Optical Microscopy (SNOM), which provides maps of the \intensdis and the topography, highlights the curvature of the wavefronts inside the lens, and the focusing.

%
%

% ---------------------------------------------------------
% --- Results and discussion}
% ---------------------------------------------------------
\section{Results and discussion}
% ---------------------------------------------------------
% --- Design and simulation
% ---------------------------------------------------------
\subsection{Design of the \hmfe}

The refractive index of the HMFE decreases from its center towards its rim following the formula\;\cite{book_bornwolf}
\begin{equation}\label{eq:index}
n(r)=\frac{n_0}{1+(r/R)^2}, %r_{max}
%\label{S3E5}
\end{equation}
where $r$ denotes the radial distance from the lens center, and $R$ denotes its radius, the latter being set as $R = 5$\mum, which scales the decrease of the refractive index. Therefore, the index is maximum at the center of the lens $n(r=0) = n_0$, and it decreases to $n(r=R) =  n_0/2$ at the rim. 
As we deal with a 220\;nm thick \ce{Si} layer set on a 2\mum\;\Sio\;substrate, we have to take into account of the effective index of the mode guided in the \Si layer, which is that of the fundamental Transverse Electric mode (\teo). It has been numerically calculated by the means of the Mode Solution Solver of Ansys Lumerical\texttrademark\;commercial software: $n_{\ce{Si}_{guided}} = 2.827$,\;\cite{Lumerical} and it is the maximum refractive index $n_0$ of the HMFE.  The minimum refractive index of the HMFE is consequently $n_{min} = 1.414$, which is lower than the refractive index of the \Sio substrate\;($n_{\Sio} = 1.444$). This poor refractive index contrast results in energy loss, because the wave is no longer perfectly guided inside the \Si\;layer, so that it penetrates the \Sio\;substrate. Nevertheless, the consecutive loss should not significantly affect the performances of the lens, because only a tiny region of the HMFE (about one external row of holes) has an effective index lower than that of the substrate $\nsio$.

Thus, the HMFE is designed through the engineering of the IFCs of PhCs. The method has been described in detail, previously.\;\cite{oc285_gaufillet, om47_gaufillet, pj8_gaufillet, pj10_gaufillet} The polarization of the incident wave is transverse electric (TE), the electric field being perpendicular to the axis
of the constitutive air holes of the PhC. 
When the lattice period of the PhC is smaller than the Bragg limit $a < \lambda/2\neff$, the PhC operates in the non-resonant metamaterial regime, and its EFCs are nearly circular. To this end, we considered hexagonal lattice PhCs, because this ensures easier access to circular IFCs. Taking into consideration the fabrication difficulty with the current nano-fabrication technology and the need that the majority of the device has to operate in the non-resonant metamaterial regime, we set the lattice constant $a$ as 290\nm. The index gradient of the HMFE is obtained by gradually varying the air filling factor $\eta$ of the PhC, which is given by\cite{yue2022dual}
\begin{equation}
\begin{aligned}
\eta=\frac{2}{\sqrt{3}}\frac{\pi \phi^2}{a^2},
\label{eq:fillingfactor}
\end{aligned}
\end{equation}
where $\phi$ denotes the air hole radius and $a$ denotes the hexagonal lattice constant. To precisely design the HMFE, we establish a calibration curve $\neff(\eta)$  in order to extrapolate the effective index of the PhC slab. To this end, we follow a two-step method as we recently demonstrated in \cite{jlt_xin} to design a Mikaelian lens.  
%
% -------------------------------------------------
% --- figure courbe calibrage
% ------------------------------------------------
\begin{figure}[H]
\centering
\includegraphics[width=\linewidth]{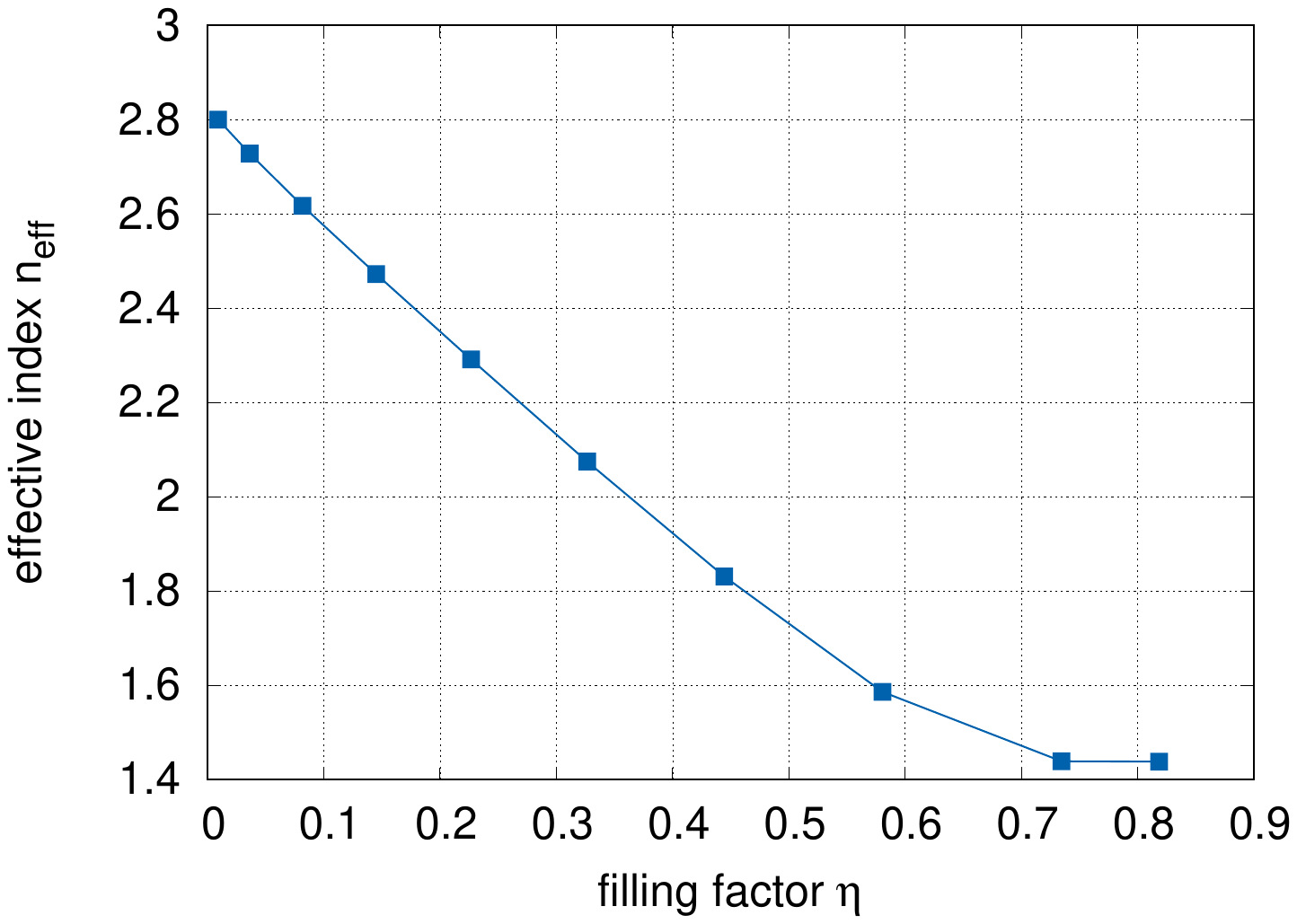}
\caption{\label{fig:calibration} Calibration curve $\neff=\neff(\eta)$ used used to design the HMFE. The effective index of the PhC slab decreases as the air filling factor $\eta$ increases. }
\end{figure}
Firstly, we calculate the effective index $\neffphc$ for an infinitely tall PhC (air hole in bulk silicon). Then, we use $\neffphc$ to replace the refractive index of the guiding layer and calculate the mode effective index of the new waveguide, denoted $\neffphcmode$.\cite{jlt_xin} 
The latter is denoted $\neff$ herein. 
The calibration curve is reported in figure\,\ref{fig:calibration}. According to it, we extrapolated the different filling factors $\eta$ in function of the radius, so as to fit the index profile of the HMFE \seeeq{eq:index}, and designed the lens. The schematic diagram of the photonic chip involving a GPC is given in Fig.\;\ref{fig:fig1b}. The air filling factor $\eta$ of the designed HMFE thus increases from the lens center toward the rim. It is zero at the center of the GhPC, and the minimum and maximum air hole radius $\phi$ of the HMFE are 23\nm ($r_{min}/a=0.08$, $\eta_{min}=0.023$) and 140\nm ($r_{max}/a=0.48$, $\eta_{max}=0.84$), respectively. These values have been chosen to enable the manufacture of the device. 

%
% ---------------------------------------------------------
% --- Simulations and results
% ---------------------------------------------------------
\subsection{Simulations of the \hmfe and results}
\begin{figure}[htbp]
\centering
\includegraphics[width=0.95\linewidth]{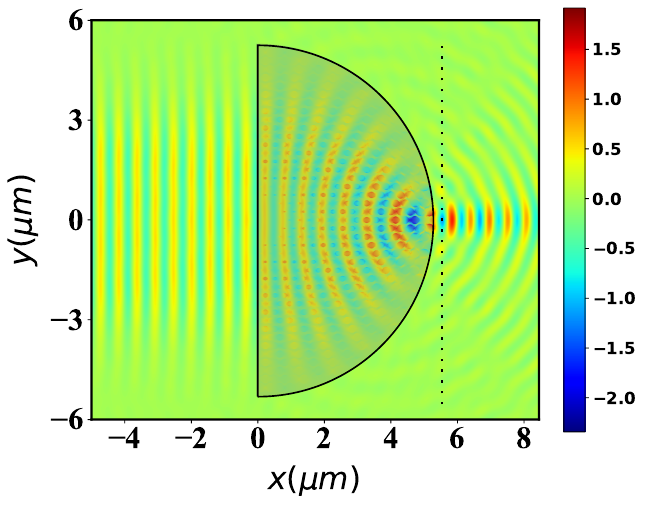}
\caption{3D FDTD simulation of the %$\instantE$ 
instantaneous $E_y$ field distribution in the symmetry \xyplane of the silicon layer at $\lbdao$=1550\nm. The HMFE is outlined by the black line. The input wave is well-focalised by the HMFE. The vertical dotted line indicates the focal plane, which is around 0.2\mum\;(0.13$\lbdao$) away from the rear-end of the HMFE.}% $x\,y$\protect\nobreakdash-plane \xyplane} %$x\,y$\nobreakdash-plane}
\label{fig:Eevolution}
\end{figure}

Next, 3D FDTD simulations of  the designed HMFE are carried out by the means of the FDTD Solution from Lumerical\texttrademark\;commercial sofware.\cite{Lumerical} The HMFE is fed by the \teo fundamental mode of a 10\mum\;wide monomode strip waveguide\;\seef{fig:fig1b}.\;\cite{acsam12_fan, apl124_yue} The simulated instantaneous $\mathbf{E}$ field distribution at $\lbdao$ in the symmetry \xyplane of the silicon layer is shown in\;\Fig{fig:Eevolution}. We observe that the wavefronts of the plane input wave are gradually curved inside the lens, and that it is eventually focused to the focal plane (see vertical dotted line in \Fig{fig:Eevolution}), which is around 0.2\mum\;(0.13$\lbdao$) far from the lens rear-end. This slight difference in the focus position is mainly ascribed to the insufficient discretisation of the index profile of the HMFE by the period lattice $a$. It may be possible to design a HMFE with a focus on the lens surface with a more precise discretisation, i.e. smaller lattice constant $a$. %Despite this slight change of the focus position, we observe a very good wave-focusing phenomenon. 
Despite this slight misalignment, it can be observed that the device effectively focuses the incident wave. %Similar results of simulation are observed for wavelengths from 1450\nm to 1550\nm???, showing the broad band of operating wavelengths \seesm{S-fig:simumultiwavelength}. 
Similar results ($\instantE$) are also observed in the 1450 to 1590\nm range of wavelengths and reported in the General Results section. \seesec{sec:3DFDTD}%  \seesm{S-fig:focalizingHMFE}.
Moreover, the HMFE should be operational for longer wavelengths, because the PhC works in the non-resonant metamaterial regime.\cite{staude2017metamaterial} %, which is not available with our experimental setup ???

\begin{figure}[htbp]
    \centering
    \begin{subfigure}[b]{0.85\textwidth} % "0.45" donne ici la largeur de l'image
        \centering 
\includegraphics[width=\textwidth]{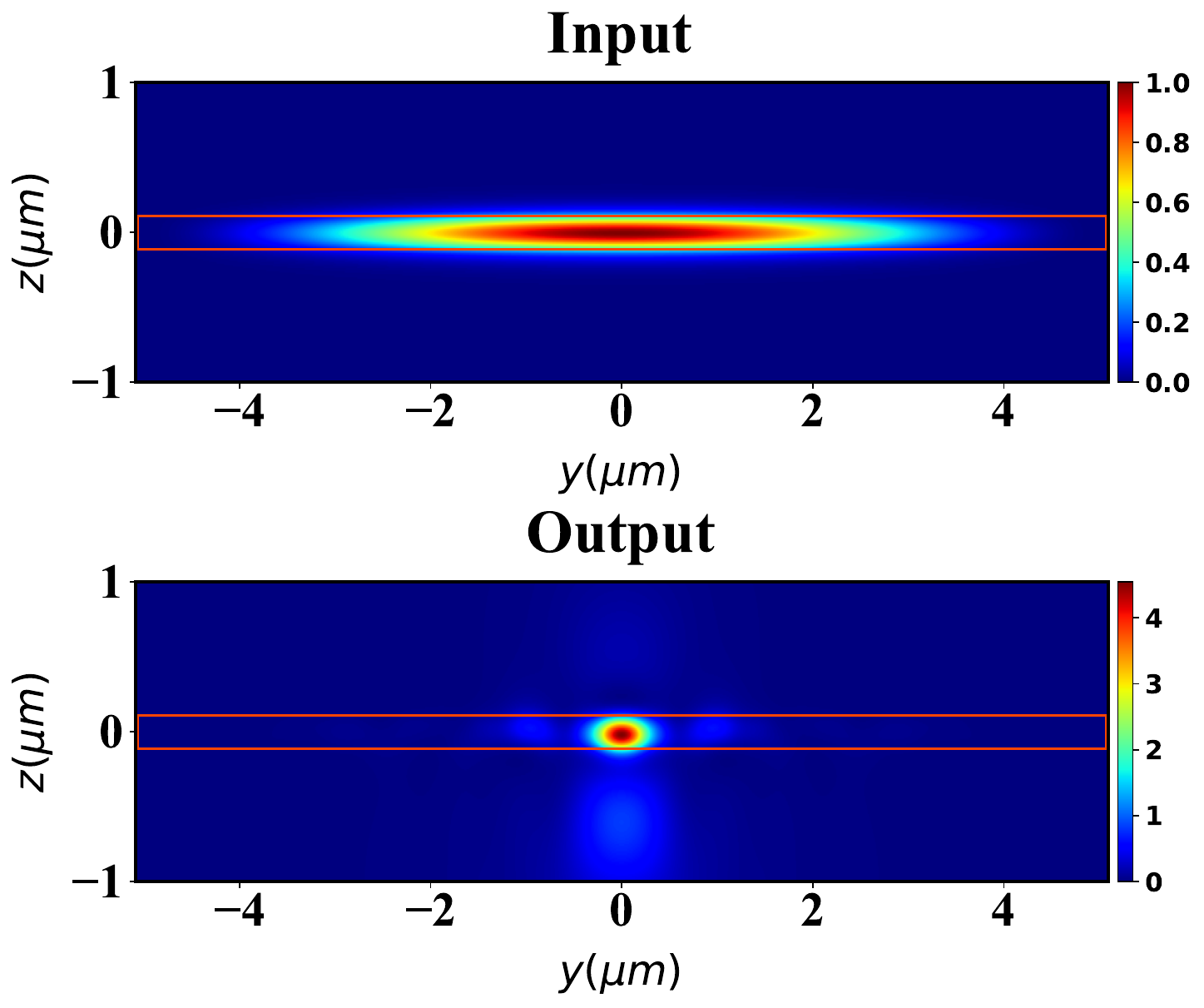}
        \caption{}\label{fig:IOHMFEa}
        
    \end{subfigure}
\hfill    ~ % ce symbole ajoute un espacement horizontal entre les premires deux images
    
 % la ligne blanche correspond au retour ˆ la ligne aprs le deuxime image   
    \begin{subfigure}[b]{0.85\textwidth}
        \centering 
        \includegraphics[width=\textwidth]{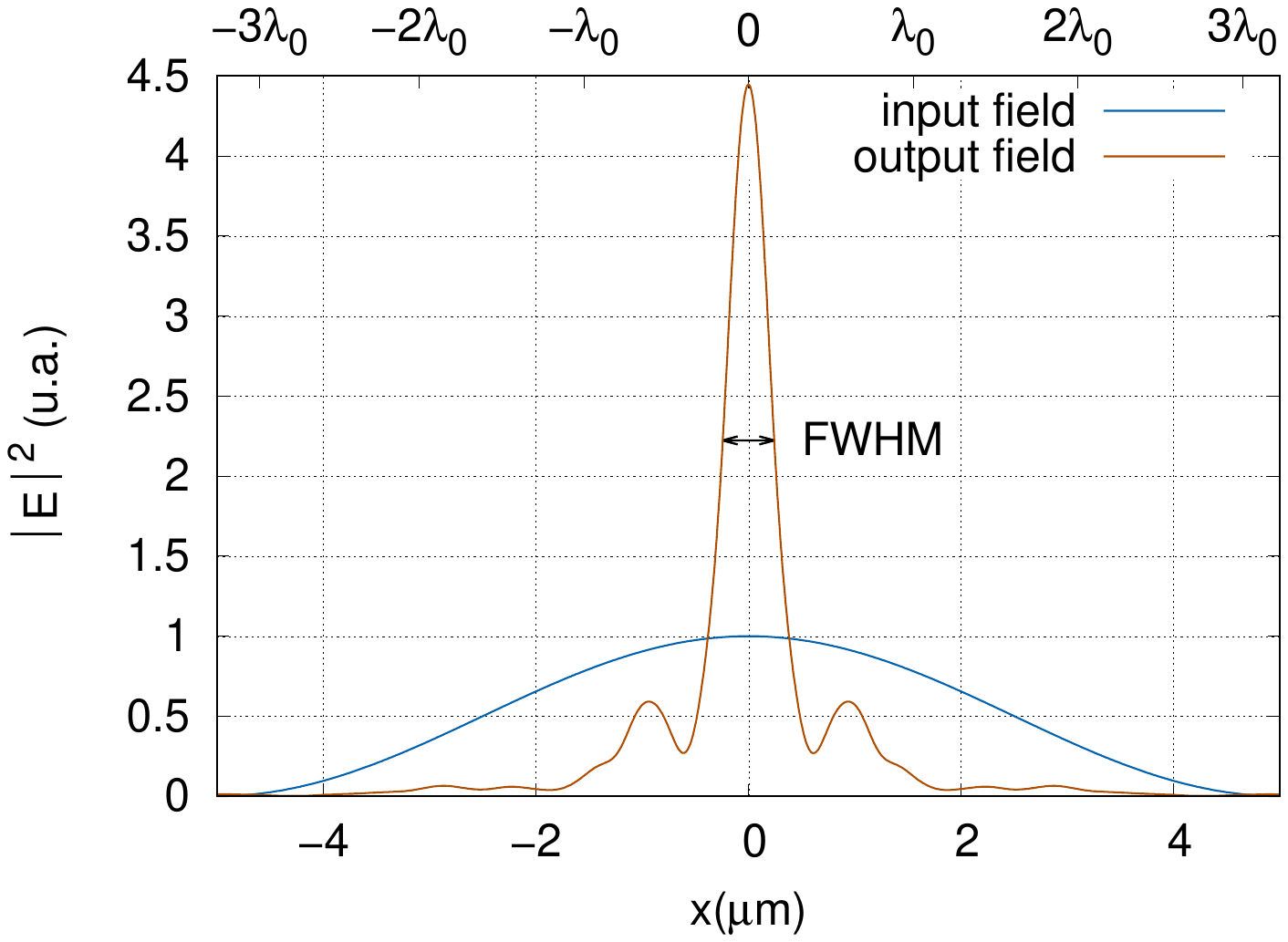}
        \caption{}\label{fig:IOHMFb}
    \end{subfigure}
    \caption{\label{fig:IOHMFE} 3D FDTD simulations of the \intensdis at $\lbdao=1.55$\mum, in two different planes of the device. (a) \Intensdis in the input and output planes (\yzplane cross-section of the chip). The two red lines delimit the \Si layer. Two secondary maxima can be observed in the output plane. It can be noticed in the bottom image that a part of the energy penetrates the \Sio substrate (below the bottom red line). (b) \Intensdis of the input and output in the symmetry (\xyplane) of the silicon guiding layer. The FWHM of the output field is 0.484\mum\;(0.312$\lbdao$).}
  \end{figure}

The \intensdis in the \yzplane at $\lbdao$ at the input and the output (along the vertical dotted line in Fig.\;\ref{fig:Eevolution}) are reported in \Fig{fig:IOHMFEa}, indicating the focusing of the incident plane wave inside the \Si layer. Nevertheless, it can be observed that the field slightly overflows in the \Sio substrate (see the bottom sub-figure of \Fig{fig:IOHMFEa}, because of the mentioned poor contrast of indexes. 
The intensity profile of the EM field in the focal plane (\xyplane) is plotted in Fig.\;\ref{fig:IOHMFb}, as well as that of the incident one. The Full Width Half Maximum (FWHM) of the output spot is 0.484\mum\;(0.312$\lbdao$). %\xyplane  \yzplane
These simulations thus indicate the excellent broadband focusing capacity of the HMFE over a wide range of wavelengths. 

% ---------------------------------------------------------
% --- Fabrication and characterisation
% ---------------------------------------------------------
\subsection{Fabrication and characterisation of the \hmfe}
% ---------------------------------------------------------
% --- Fabrication
% ---------------------------------------------------------
\subsubsection{Fabrication}

\begin{figure}
\centering
\begin{subfigure}{0.995\textwidth}
    \includegraphics[width=\textwidth]{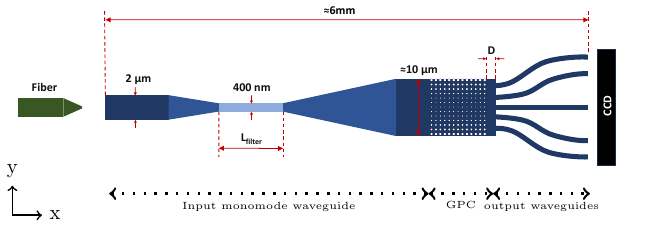}
    \caption{Sketch of the overall structure with the fan-shaped set of output waveguiedes designed for characterization with the CCD camera (unscaled). It consists of three parts: (i) the designed input single mode waveguide; (ii) the GPC; (iii) the output fan-shaped set of waveguides. D is the distance between the output interface of the lens and the fan-shaped set of output waveguides.}
    \label{fig:structure_fanshaped}
\end{subfigure}
%\hfill ~

\begin{subfigure}{0.995\textwidth}
    \includegraphics[width=\textwidth]{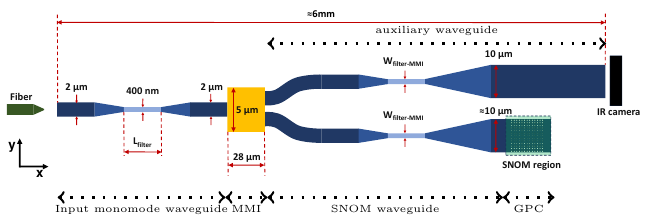}
    \caption{Sketch of the overall structure designed for the characterization by SNOM (unscaled). It consists of five parts: (i) the designed input single mode waveguide;
(ii) the 1$\to$ 2 multimode interference (MMI) splitter; (ii) the GPC; (iv) the auxiliary
waveguide for alignment; (v) the arm for SNOM characterization with the GPC.}
    \label{fig:structure_snom}
\end{subfigure}
\caption{The two types of structures used for characterization. (a) with the fan-shaped set of output waveguides. (b) for SNOM characterization. }
\label{fig:structures}
\end{figure}

From the simulation results, the HMFE is fabricated on a SOI wafer made of a 220\nm\;thick silicon (\Si) layer set on a 2\mum\;thick buried oxide (\Sio) layer. Two structures have been fabricated, one for each method of characterization, sketches of which are shown in figure\;\ref{fig:structures}.
The HMFE is fed by a 10\mum wide strip Silicon waveguide. As is shown in the optical microscopy and Scanning Electron Microscope (SEM) images \seef{fig:SEMHMFE}, the HMFE is generally successfully fabricated with clean PhC structures except that some ultra-tiny holes are not perfectly etched. Its diameter is 10.3\mum (from the leftmost hole edge to the rightmost hole edge). 
If the finest holes could not be cleanly etched (diameter = 46\nm according to the design), it is noticeable that the diameter of the smallest etched holes is 54\nm \seef{fig:semf}. 
Since these ultra-tiny uncorrectly etched holes can only lead to a very small effective index change, they should have very little impact on the functionality of the HMFE. The distance between the input waveguide and the HMFE is set as 5\mum\;so as to avoid the proximity effect during the nano-fabrication process.\cite{acsam12_fan, apl124_yue} The details about the device fabrication are presented in the Methods section.\;\seesec{sec:device_fabrication}

\begin{figure}%{width=0.8\textwidth}
\centering
\begin{subfigure}{0.475\textwidth}
    \includegraphics[width=\textwidth]{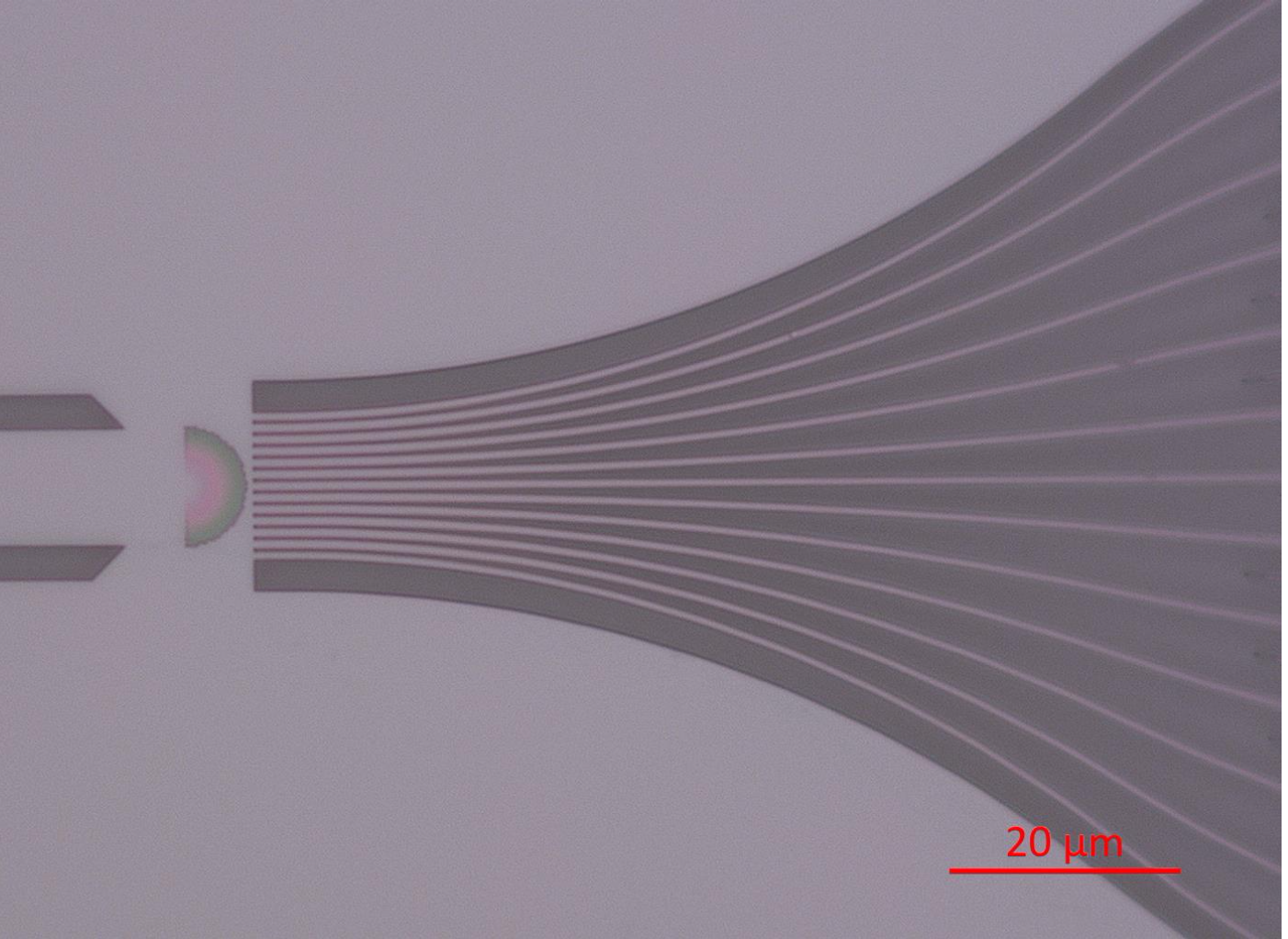}
    \caption{}
    \label{fig:sema}
\end{subfigure}
\hfill
\begin{subfigure}{0.475\textwidth}
    \includegraphics[width=\textwidth]{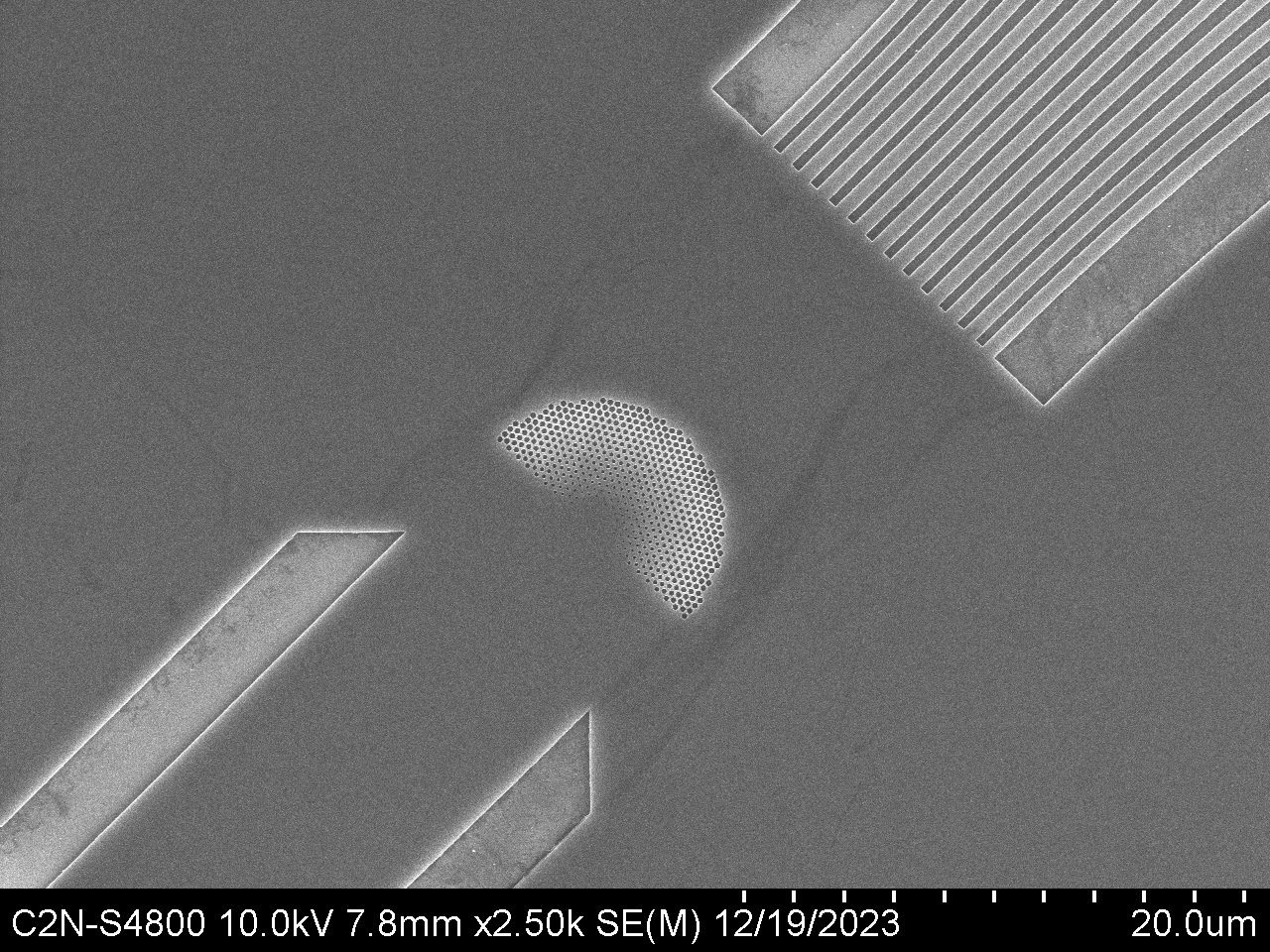}
    \caption{}
    \label{fig:semb}
\end{subfigure}
\hfill
\begin{subfigure}{0.475\textwidth}
    \includegraphics[width=\textwidth]{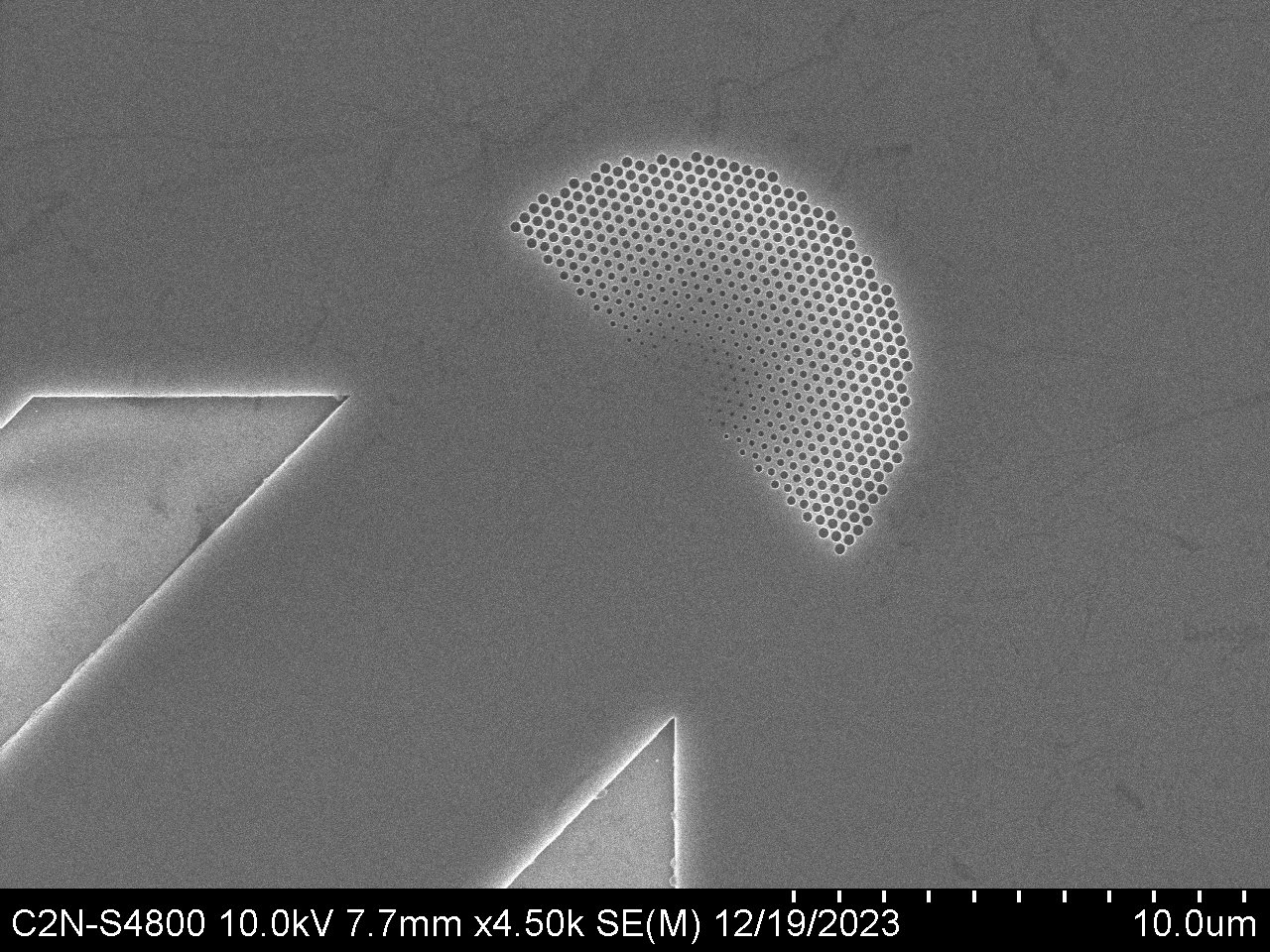}
    \caption{}
    \label{fig:semc}
\end{subfigure}
        \hfill
\begin{subfigure}{0.475\textwidth}
    \includegraphics[width=\textwidth]{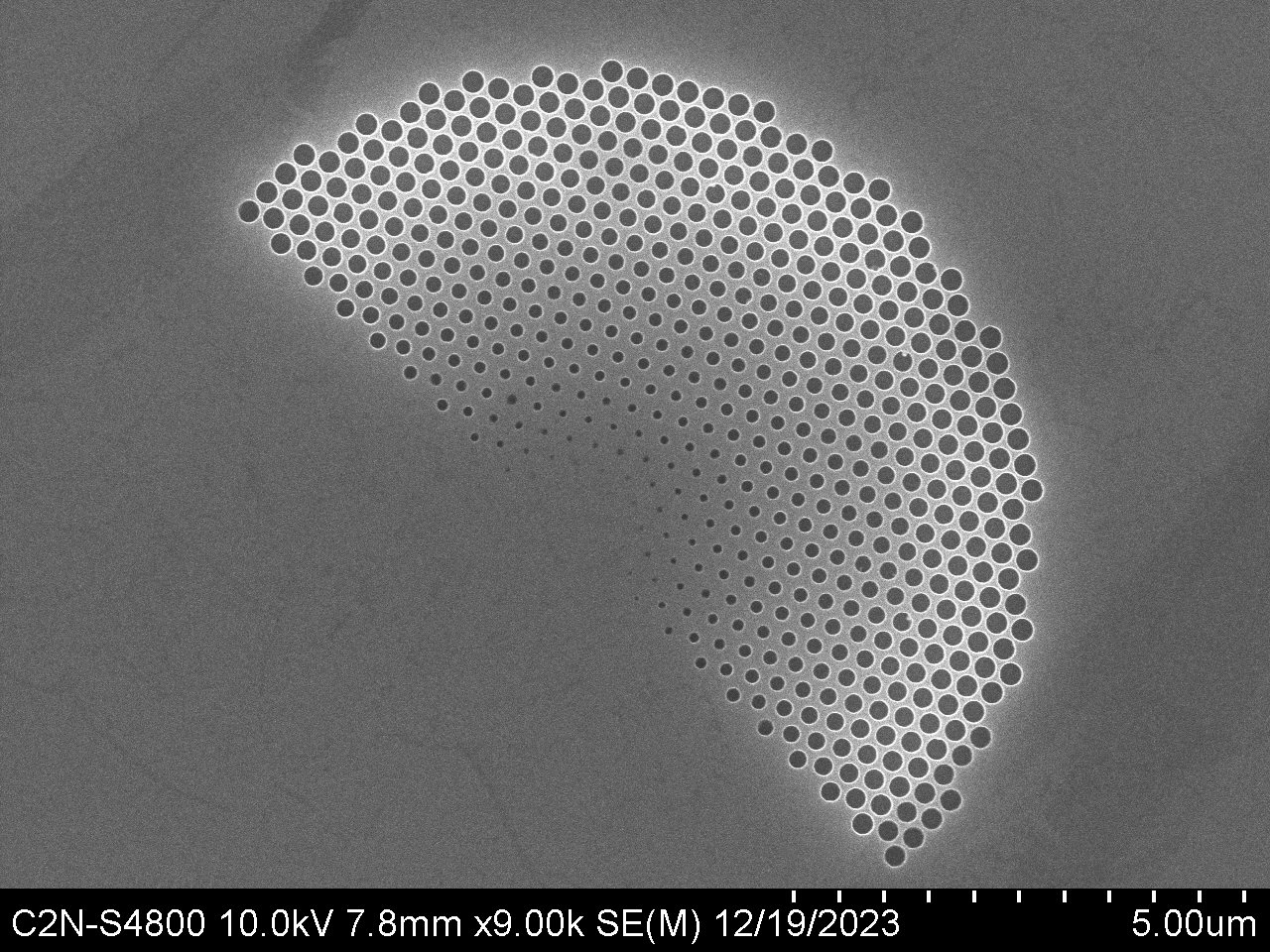}
    \caption{}
    \label{fig:semd}
\end{subfigure}
         \hfill
\begin{subfigure}{0.475\textwidth}
    \includegraphics[width=\textwidth]{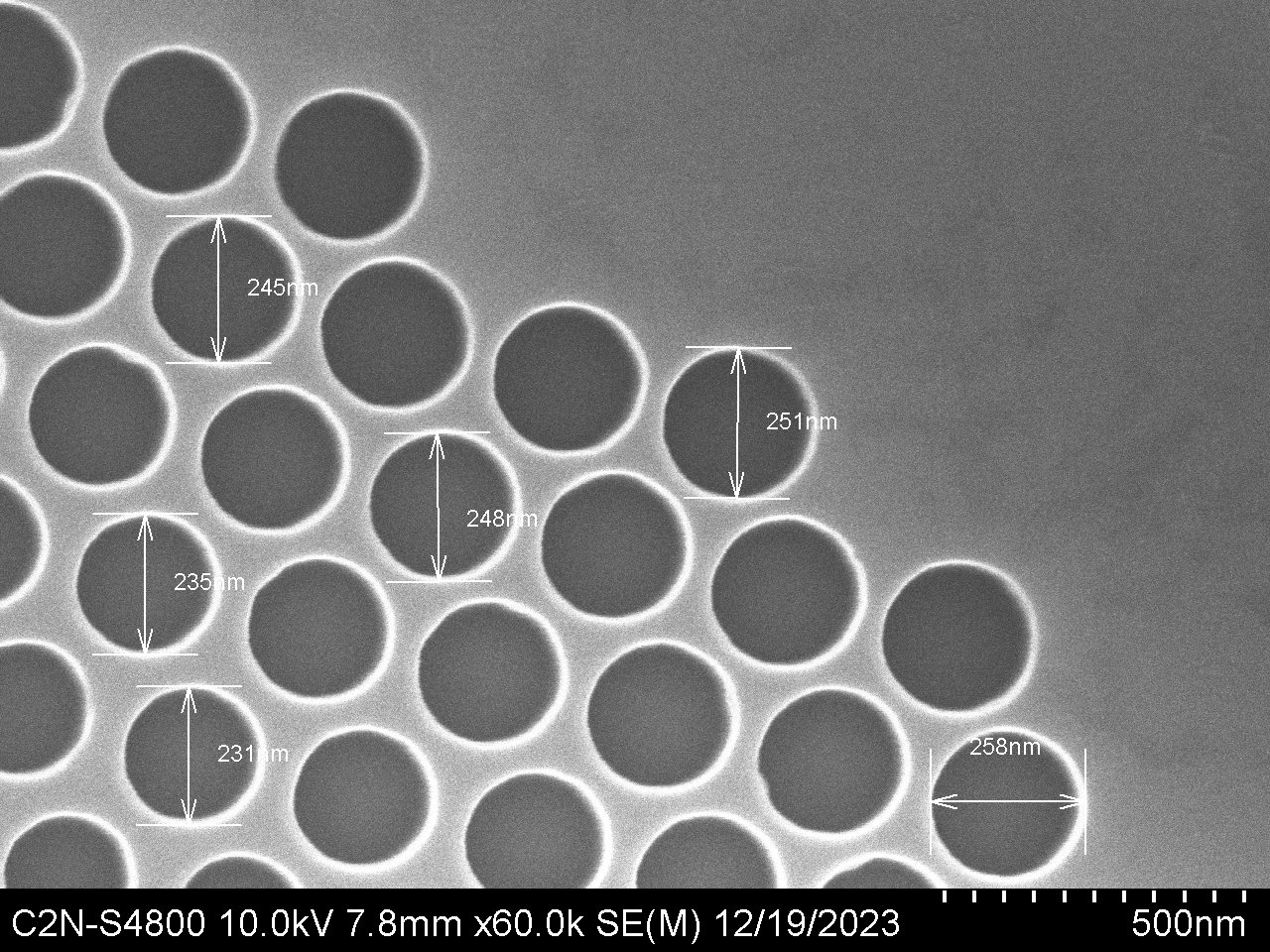}
    \caption{}
    \label{fig:seme}
\end{subfigure}
        \hfill
\begin{subfigure}{0.475\textwidth}
    \includegraphics[width=\textwidth]{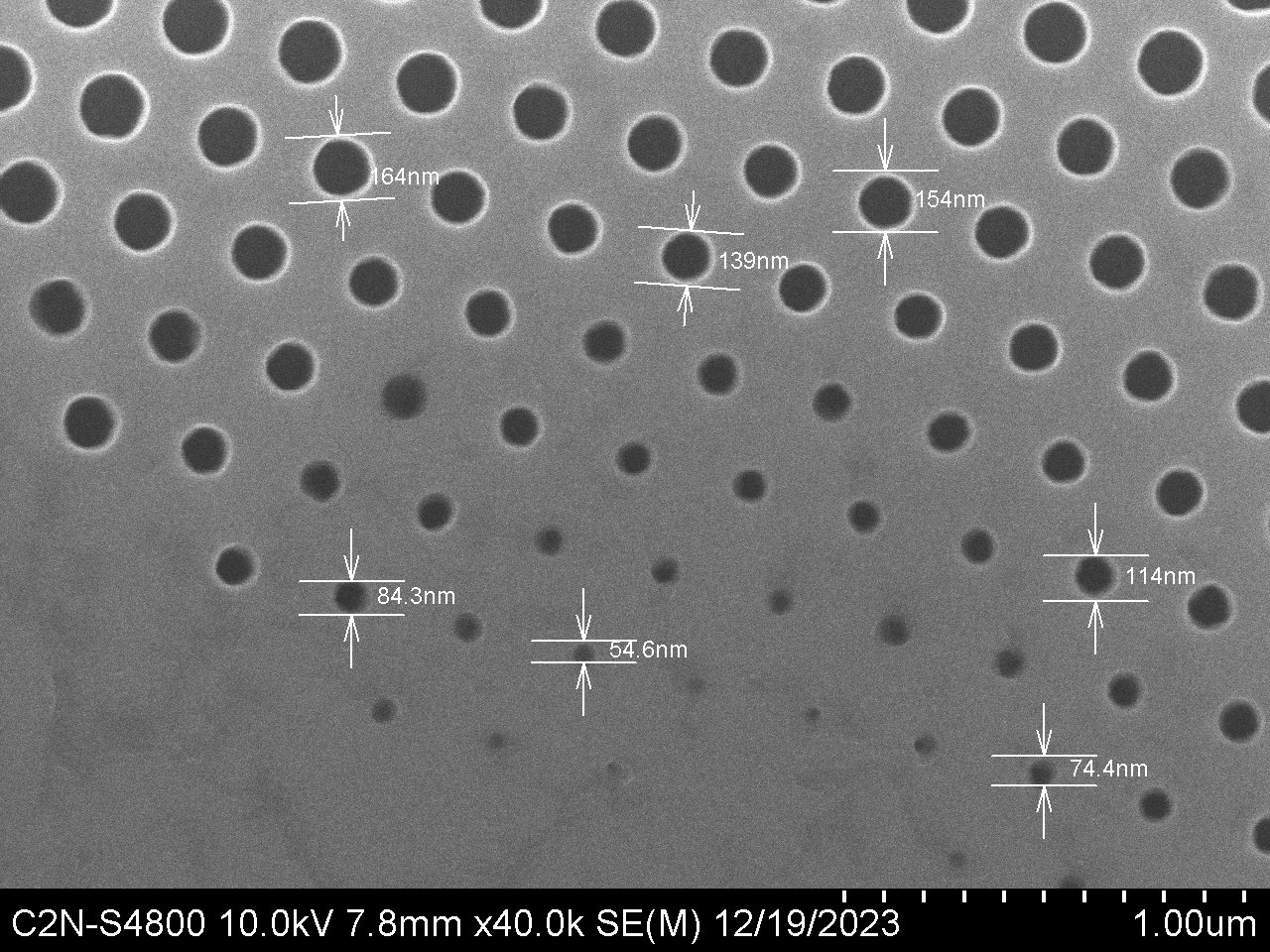}
    \caption{}
    \label{fig:semf}
\end{subfigure}
\singlespace\caption{Optical microscopy (a) and SEM images (b-f) of the two fabricated devices. (a) Optical microscopy image of the fan-shaped output waveguides region (global view): it shows the input strip waveguide, the GPC and the thirteen fan-shaped output waveguides, which are all etched in the Silicon slab. (b) Input waveguide that is incident onto the GPC, plus the output fan-shaped set of thirteen waveguides. The input waveguide and the GPC are 5\mum apart. SEM image of the fan-shaped output waveguides region (zoomed view). 
(c) SEM image of the structure for the SNOM characterisation: input waveguide and HMFE, which are also 5\mum apart. (d) SEM image of the HMFE, showing the radially increasing air filling factor. (e) Zoomed view of the circular border (the rim) of the HMFE. (f) Zoomed view of the center of the HMFE.}
\label{fig:SEMHMFE}
\end{figure}

% ---------------------------------------------------------
% --- Characterisation with fan-shaped output waveguides
% ---------------------------------------------------------
\subsubsection{Characterisation with the fan-shaped set of output waveguides}
\begin{figure}[H]
\centering
\begin{subfigure}{0.85\textwidth}
    \includegraphics[width=\textwidth]{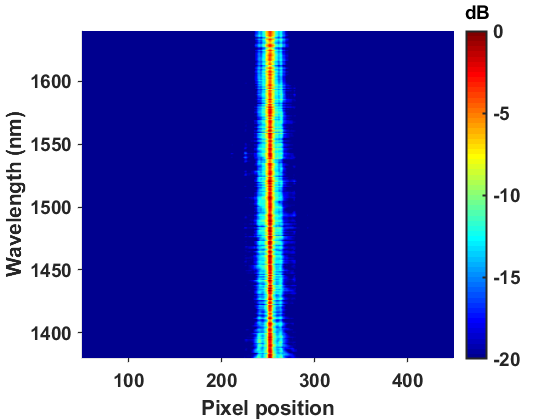}
    \caption{}
    \label{fig:colormap_fanshapeda}
\end{subfigure}
\hfill

\begin{subfigure}{0.85\textwidth}
    \includegraphics[width=\textwidth]{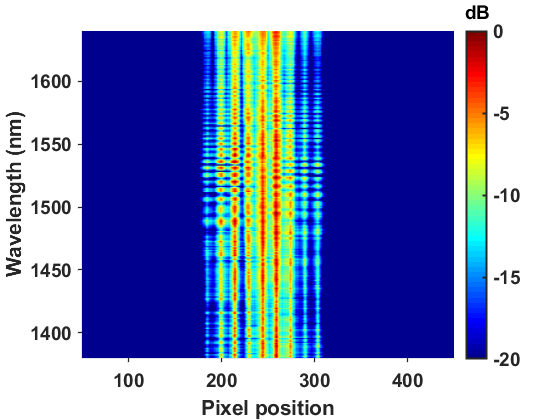}
    \caption{}
    \label{fig:colormap_fanshapedb}
\end{subfigure}
       
\caption{Colormap of the \intensdis into the fan-shaped set of output waveguides as a function of the wavelengths (1380\nm to 1640\nm) (a) for the structure with the HMFE as $D = 0$\mum. (b) for the reference structure (without the HMFE).}
\label{fig:colormap_fanshaped}
\end{figure}

Firstly, a set of fan-shaped output waveguides is used to characterise the fabricated HMFE \seefd{fig:sema}{fig:semb}. This method has been firstly exposed in \cite{oe12_lupu} and also involved in \cite{apl124_yue}, and %the experience on a plasmonic lens \cite{fan2017integrated} and a PhC super-prism.\cite{lupu2004experimental} + flat lens 
efforts have been devoted to the design of the input waveguide structure to filter high-order $\mathbb{TE}$ modes, so that only the fundamental \teo mode ideally propagates in the waveguide. The input wave thus passes through the HMFE and is then collected by thirteen fan-shaped 600\nm wide waveguides, spaced 400\nm apart, and followed by a CCD camera. The separation distance between these output waveguides increases rapidly in the fan-shaped region so as to avoid energy coupling between them. %By varying the distance $D$ between the rear end of the lens and the fan-shaped output waveguides, we can obtain the field intensity distribution at different positions. 
The results in the [1380:1640]\nm range are reported in %\ref{fig:opticalcarac}(a) 
\Fig{fig:colormap_fanshapeda} as  $D=0$\mum, and it can be noticed that the output wave is mainly concentrated in the central waveguide after progagating through the HMFE\;%\seef{fig:opticalcarac}(a). 
\seef{fig:colormap_fanshapeda}. 
%The distance $D$ has been thus varied from $D_{min}$ = 0.5\,$\mu$m to $D_{max}$ = 30\mum\; with a step $\Delta D = 0.5\;\mu$m.
%D=0, D=0.5, D=1,D=1.5...jusqu'ˆ D=14.5 µm avec un step de 0.5 µm 
The distance $D$ has been also varied from $D_{min}$ = 0\,$\mu$m to $D_{max}$ = 14.5\mum, with a step $\Delta D = 0.5\;\mu$m, which allows to determinate the focal length. Thirty structures have been thus fabricated, with the same lens but varied distance $D$. 
%
%($D = 1, 2$ and 3\mum) 
%
When $D$ increases, the output wave spreads over several waveguides, because the output wave begins to diverge, % \seesm{S-fig:focalizingHMFE}, 
which can be seen in the General Results section \seesec{sec:fanshaped_d}. This indicates that the focal length is $D = 0$\mum. A reference structure (without the lens) and consisting of nine fan-shaped 600\nm\;wide waveguides) is also characterised for comparison\;%seef{fig:opticalcarac}(b) 
\seef{fig:colormap_fanshapedb}. The output wave is found to be, in that case, spread over all the nine waveguides of the reference structure, so we can conclude to very good focusing capability of the device in this wide range of wavelengths. Details about the experimental setup are given in the following Methods section\;\seesec{fanshapedmethod}.
%Therefore, we can conclude a  by comparing it with the result of a reference structure. 

% ---------------------------------------------------------
% --- SNOM characterisation
% ---------------------------------------------------------
\subsubsection{SNOM characterisation}
Secondly, the device is characterized by Scanning Near-Field Optical Microscopy (SNOM) by the means of an end-fire coupling system. %
The SNOM probe is moved above and very close to surface of the HMFE ($\approx$ 10\;nm) to scan it, collecting the evanescent waves, so as to provides 2D maps of the \eintensdis. \cite{lee2018anderson,vo2012near,fabre2008optical, aplp10_briche, oe33_kmiche} 

\begin{figure}[H]
\centering
\begin{subfigure}{0.85\textwidth}
    \includegraphics[width=\textwidth]{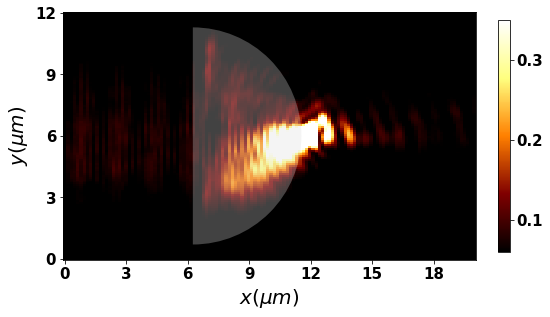}
    \caption{}
    \label{fig:snom_measa}
\end{subfigure}
\hfill

\begin{subfigure}{0.85\textwidth}
    \includegraphics[width=\textwidth, trim = 0cm 2cm 0cm 2cm, clip=true]{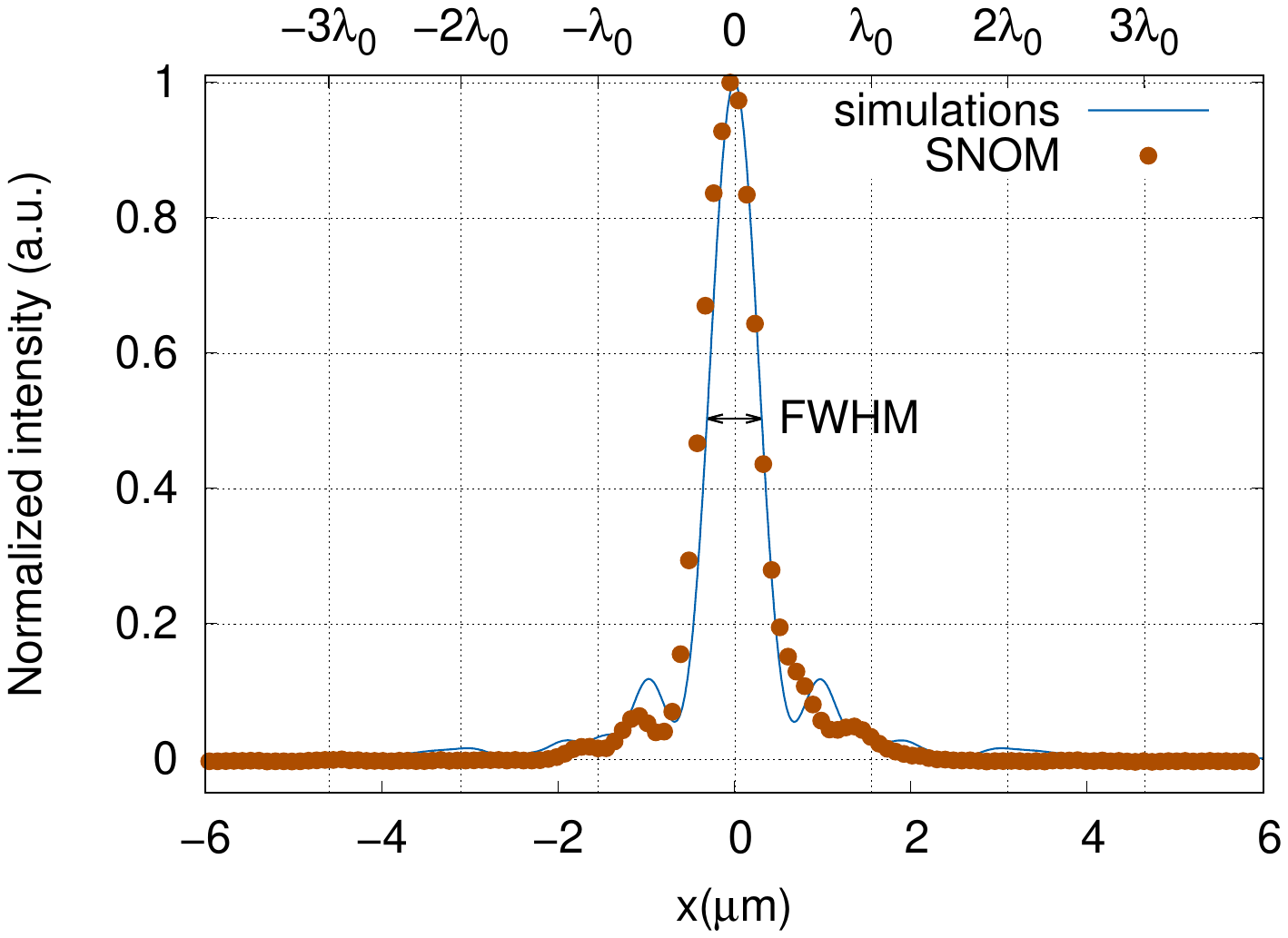}
    \caption{}
    \label{fig:snom_measb}
\end{subfigure}
       
\caption{SNOM characterization of the HMFE. (a) The SNOM image of the HMFE at $\lbdao$=1550\nm. The grey area represents the HMFE. The input wave is well-focalised, and clear wavefront curvature inside the HMFE is observed. (b) Comparison of the SNOM results (brown dots) with the simulated $\intens$ field at the rear-end of the lens (blue solid line).}% $D = $ ???}
\label{fig:snom_meas}
\end{figure}

The SNOM image of the HMFE lens at $\lbdao$=1550\nm is shown in% \Fig{fig:SNOMHMFE}(a)
\;\Fig{fig:snom_measa}. The grey area represents the HMFE. We can see that the incident wavefront is gradually curved inside the HMFE due to the index gradient and that the input wave is focalised into a bright spot right next to the rim of the lens. It is a convincing evidence of the focalising capacity of the fabricated HMFE. Its behaviour is in good agreement with the simulation results. Besides, we also observe a higher field intensity at the bottom side of the structure. We ascribe this to the insufficient filtering of the \teu mode of the input waveguide, which leads to an asymmetrical electric incident field distribution. Next, we extract the \eintensdis profile on the rear-end of the lens, for a more quantitative analysis. The SNOM measurements (brown dots) are compared with the 3D FDTD simulated \eintensdis field at the rear-end of the lens surface (blue solid line), as is shown in%\Fig{fig:SNOMHMFE}(b) 
\;\Fig{fig:snom_measb}. The SNOM result perfectly agrees with the simulation result. Next, we fitted these measured SNOM data in the focal plane with a Gaussian fit, so as to get the value of its Full Width Half Maximum (FWHM), that is, FWHM = 0.723\mum\;(0.466 $\lbdao$), which is very close to the FWHM obtained by the simulation FWHM = 0.624\mum\;(0.403 $\lbdao$).

%ioprofile.pdf

The SNOM characterization is repeated varying the input wavelength from 1450\nm to 1590\nm, with a step  $\Delta \lambda = 20$\nm, providing 2D maps of the \eintensdis inside the device, which is reported in the General Results section \seesec{sec:snom_wavelength}. % \seesm{S-fig:SNOMmultiwave}. % as $\lambda$ =  1450\nm ,  1470\nm ,  1490\nm ,  1510\nm ,  1530\nm ,  1550\nm ,  1570\nm , and  1590\nm \seesm{S-fig:SNOMmultiwave}. 
The corresponding intensity profiles are then extracted and reported in the General Results section \seesec{sec:fwhmsnom_lambda}. % the supplementary material \seesm{S-fig:F4SNOMmultiwave}. 
One may conclude that the HMFE focuses the incident wave in this broad band of wavelengths.
These extracted FWHM are next compared with the FWHM from the simulation and fitted with a linear fit\;\seef{fig:fwhmsnom}. The good agreement between the two demonstrates the focusing performances of the device.
The SNOM experimental setup is described below in the Methods section\;\seesec{sec:snom_method}.

\begin{figure}[htbp]
\centering
\includegraphics[width=\linewidth]{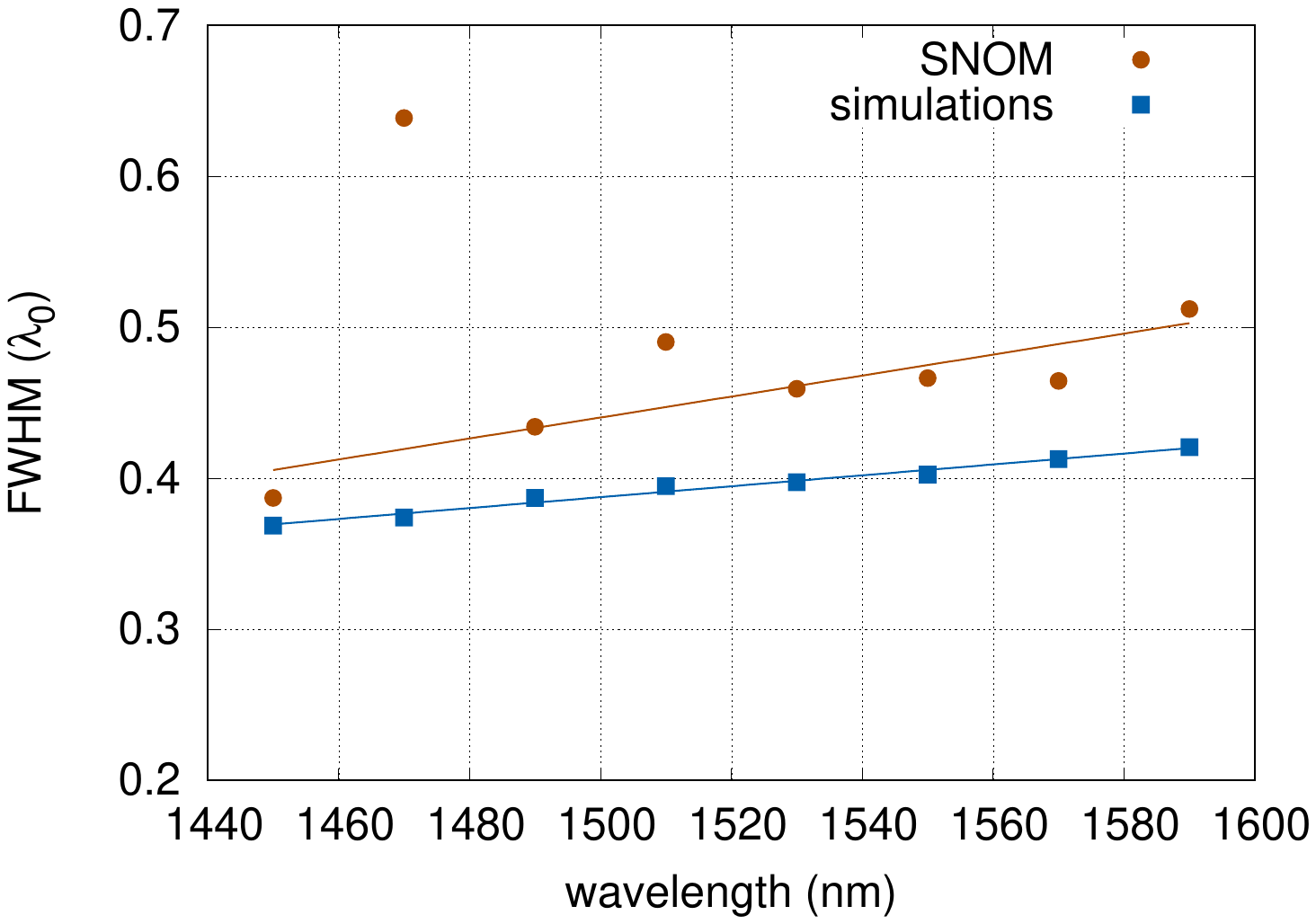}
\caption{The FWHM of the SNOM measurements (brown dots) and FWHM of the simulation results (blue dots) as a function of the wavelength. The linear fits of the data are shown in solid lines (It does not include the measurement at 1470\nm).}
\label{fig:fwhmsnom}
\end{figure}

% ---------------------------------------------------------
% --- Conclusion
% ---------------------------------------------------------
\section{Conclusion}
In this article, we proposed the design of a Half Maxwell Fish-Eye \via GRaded INdex optics. It is based on a Graded Photonic Crystal for Silicon Photonics, and we experimentally evidenced its focalising ability at 1.55\mum. We firstly numerically studied its performance across a wide wavelength range. Then, these were experimentally confirmed in two steps in the [1450:1590]\nm range, firstly by the means of a fan-shaped set of output waveguides. A further SNOM characterisation highlighted the curvature of the wavefronts inside the lens, and the focusing. The FWHM of the focus intensity profile (FWHM = $0.466\lbdao$ at $\lbdao$=1550\nm) recorded using SNOM perfectly matches with the simulated FWHM, highlighting the functionality. Because, it operates in the non-resonant metamaterial regime, the HMFE should be operational for longer wavelengths. 
Our results confirm that Graded INdex optics is the way for Photonic Integrated Circuits in Silicon Photonics.

%

% ---------------------------------------------------------
% --- General results 
% ---------------------------------------------------------
\section{General results}\label{sec:general_results_section}
% ---------------------------------------------------------
% --- 3D TDTD
% ---------------------------------------------------------
\subsection{3D FDTD simulated $\instantE$ field evolution of HMFE at different wavelengths: 1.45\mum to 1.59\mum.}\label{sec:3DFDTD}% (from $1.45\ \mu m$ to $1.59\ \mu m$).}
\begin{figure}[H]
\centering
\includegraphics[width=0.8\linewidth]{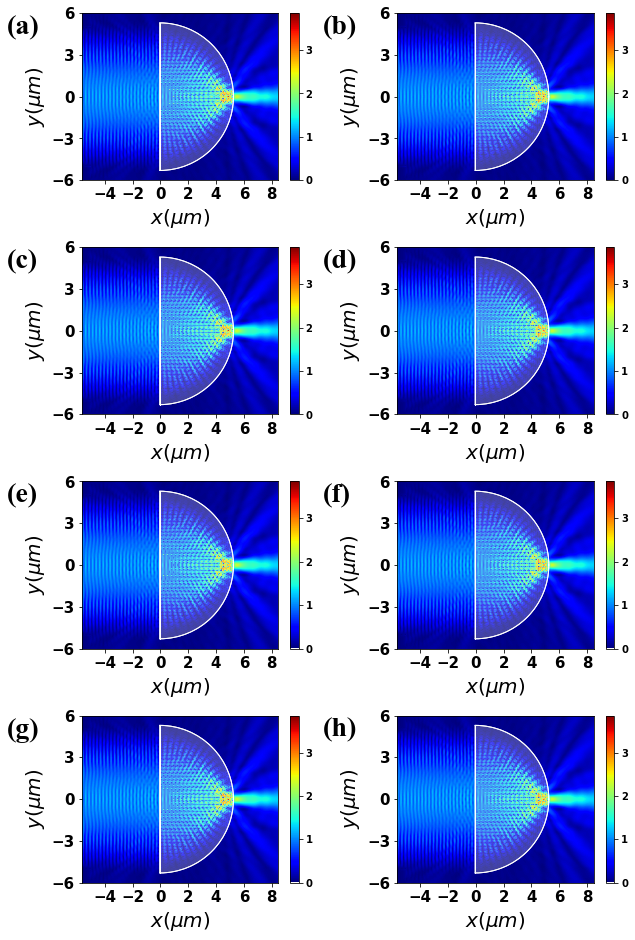}
\caption{3D FDTD simulation of the field intensity distribution $\instantE$ in the symmetry \xyplane of the silicon layer at different wavelengths: (a) $1.45$\mum, (b) $1.47$\mum, (c) $1.49$\mum, (d) $1.51$\mum, (e) $1.53$\mum, (f) $1.55$\mum, (g) $1.57$\mum, and (h) $1.59$\mum. The grey area represents the HMFE.}
\label{fig:simumultiwavelength}
\end{figure}

% -------------------------------------------------------------------
% --- Characterisation of the HMFE using fan-shaped output
% -------------------------------------------------------------------
\subsection{Characterisation of the HMFE using fan-shaped output waveguides, varying the distance $D$ between the rim of the HMFE and the fan-shaped set of output waveguides.}\label{sec:fanshaped_d}

\begin{figure}[H]
\centering
\includegraphics[width=1.2\linewidth]{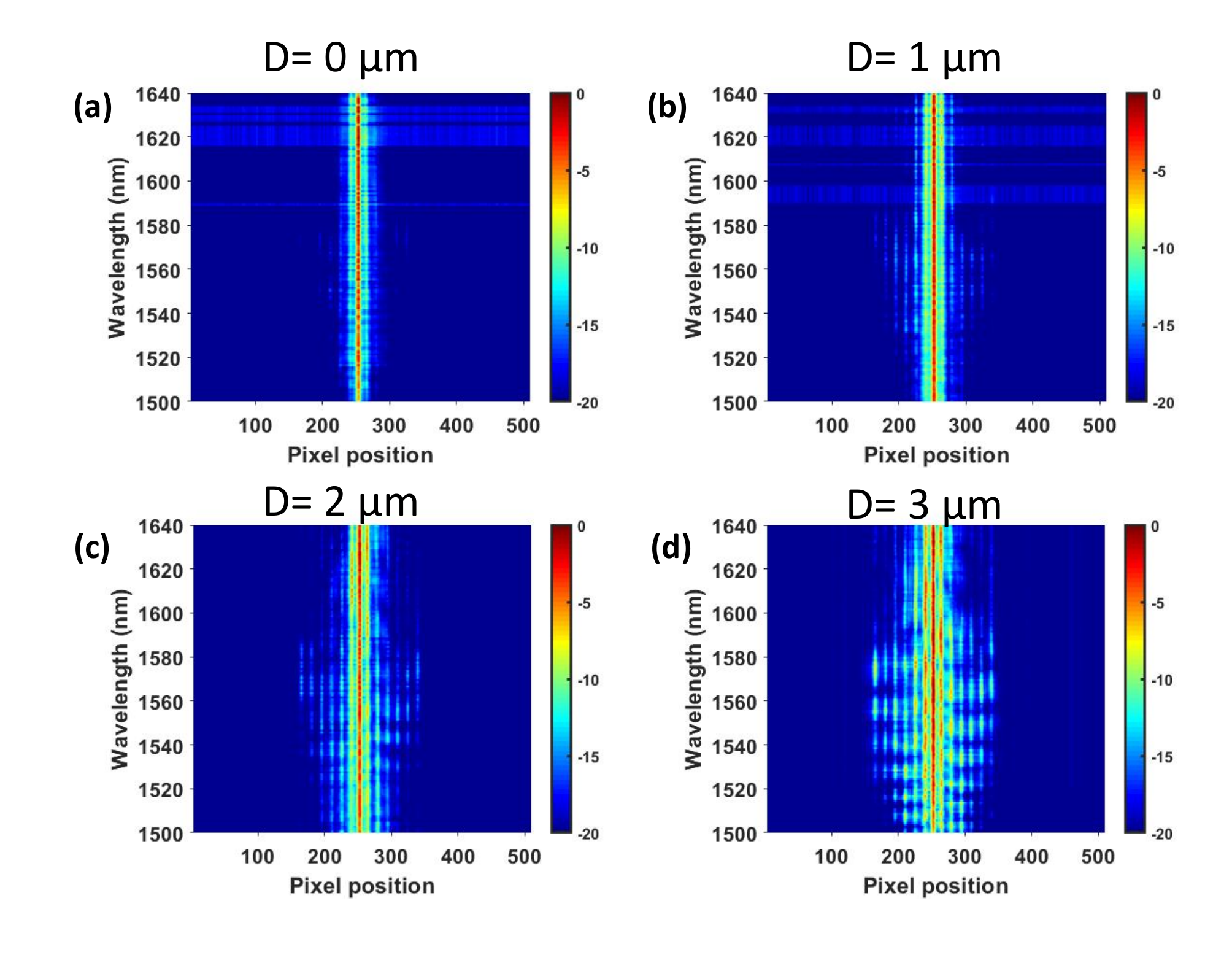}
\caption{Colormap of the \intensdis in function of the wavelength as the distance $D$, between the rim of the HMFE and the he fan-shaped set of output waveguides, is (a) $D=0$\mum, (b) $D=1$\mum, (c) $D=2$\mum and (d) $D=3$\mum. }
\label{fig:focalizingHMFE}
\end{figure}

% -------------------------------------------------------------------
% --- SNOM images of the HMFE
% -------------------------------------------------------------------
\subsection{The SNOM images of the HMFE region at different wavelengths (from $1.45$\mum to $1.59$\mum).}\label{sec:snom_wavelength}

\begin{figure}[H]
\centering
\includegraphics[width=0.9\linewidth]{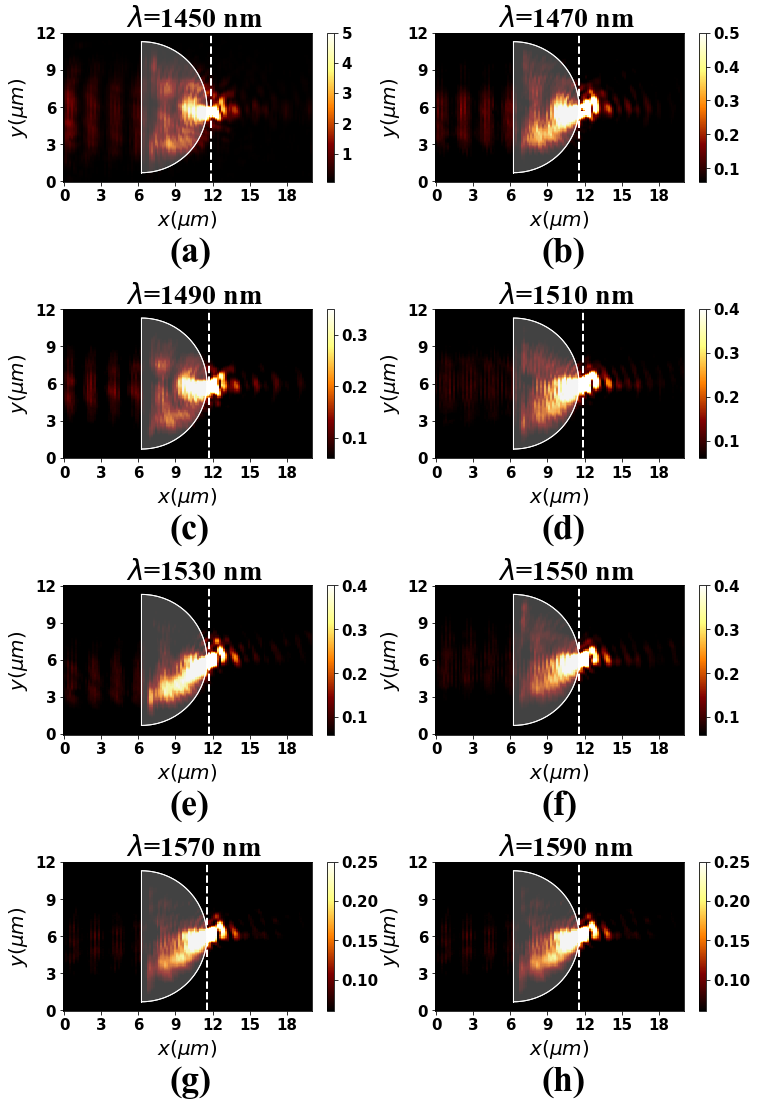}
\caption{The SNOM images of the HMFE region at $\lambda=1450\ nm$ (a), $\lambda=1470\ nm$ (b), $\lambda=1490\ nm$ (c), $\lambda=1510\ nm$ (d), $\lambda=1530\ nm$ (e), $\lambda=1550\ nm$ (f), $\lambda=1570\ nm$ (g), and $\lambda=1590\ nm$ (h). The grey area corresponds to the HMFE and the white dashed line represents the focus point position. }
\label{fig:SNOMmultiwave}
\end{figure}

% -------------------------------------------------------------------
% ---  Experimental intensity and simulated distribution profiles
% -------------------------------------------------------------------
\subsection{Experimental intensity and simulated ($\intens$) distribution profiles in the focal plane from SNOM characterisation from 1.45\mum to 1.59\mum.}\label{sec:fwhmsnom_lambda}
These experimental profiles are extracted from \Fig{fig:SNOMmultiwave}. The corresponding FWHM are reported in \Fig{fig:fwhmsnom}, and compared with those of the simulation. %\;\seef{fig:fwhmsnom}.
\begin{figure}[H]
\centering
\includegraphics[width=0.75\linewidth]{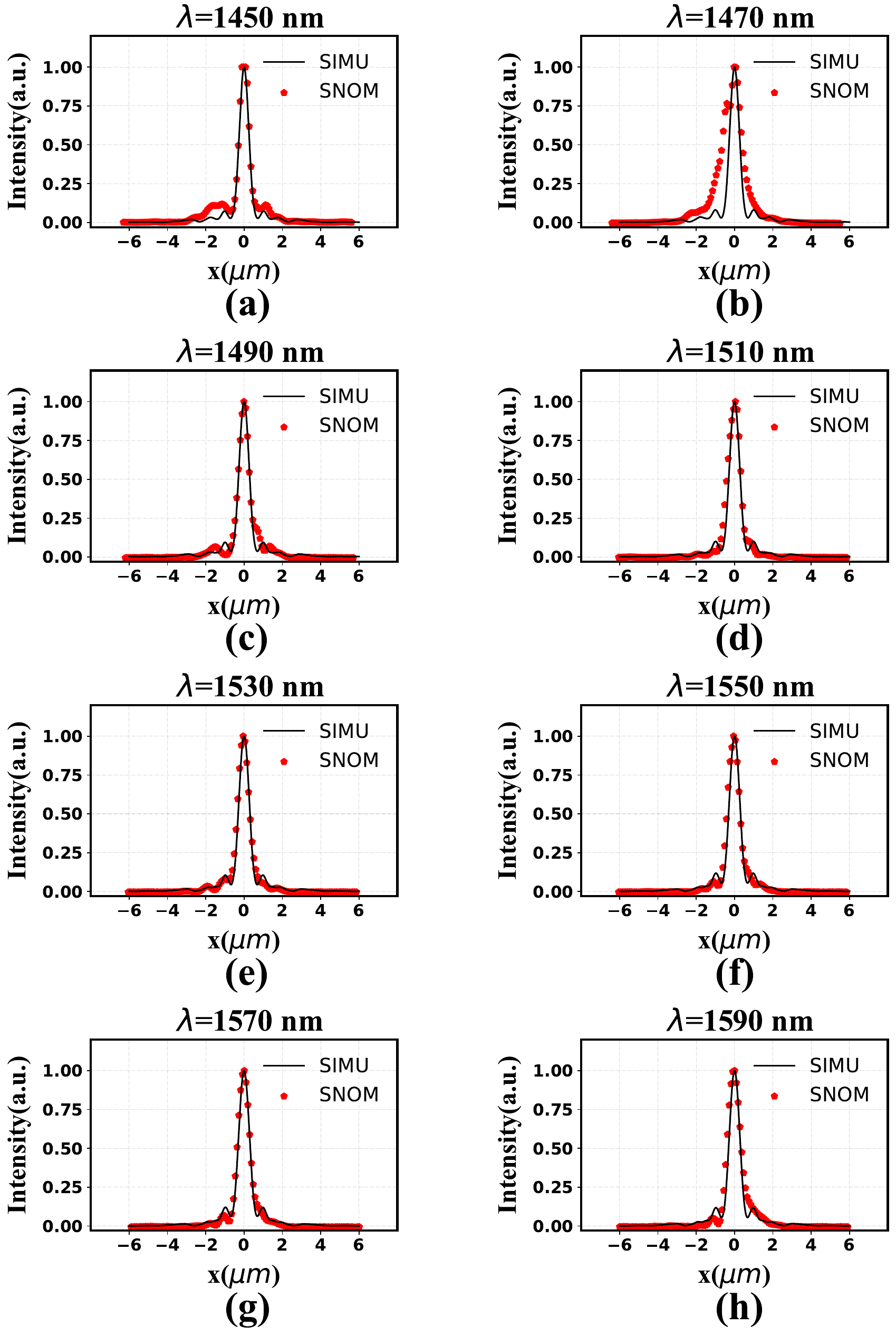}
\caption{The experimental focus intensity distribution profile is plotted with the $\lvert E \rvert^2$ distribution of the focus point in the simulation. The SNOM characterisation results agree well with the simulation result for different wavelengths.}
\label{fig:F4SNOMmultiwave}
\end{figure}

% ---------------------------------------------------------
% --- Methods section
% ---------------------------------------------------------
\section{Methods}\label{sec:method_section}
% ---------------------------------------------------------
% --- Device fabrication
% ---------------------------------------------------------
\subsection{Device fabrication}\label{sec:device_fabrication}
An undoped SOI wafer from SOITEC (220\nm\;thick silicon layer set on 2\mum\;thick buried oxide layer) is used for the fabrication of the HMFE lens. The structure definition is performed using electron beam lithography (EPBG5200) using Csar 62 positive resist. The remaining resist after development serves as a mask for the silicon etching. Then, the pattern is transferred to the silicon layer by the means of Inductively Coupled Plasma (ICP) etching, with a \ce{SF6}/\ce{C4F8} gas mixture (see also\;\cite{apl124_yue}). 

% ---------------------------------------------------------
% --- Characterisation with output waveguides
% ---------------------------------------------------------
\subsection{Characterisation with the fan-shaped set of output waveguides}\label{fanshapedmethod}
\begin{figure}[htbp]
    \centering
    \begin{subfigure}[c]{0.95\textwidth} % "0.45" donne ici la largeur de l'image
        \centering \includegraphics[width=\textwidth]{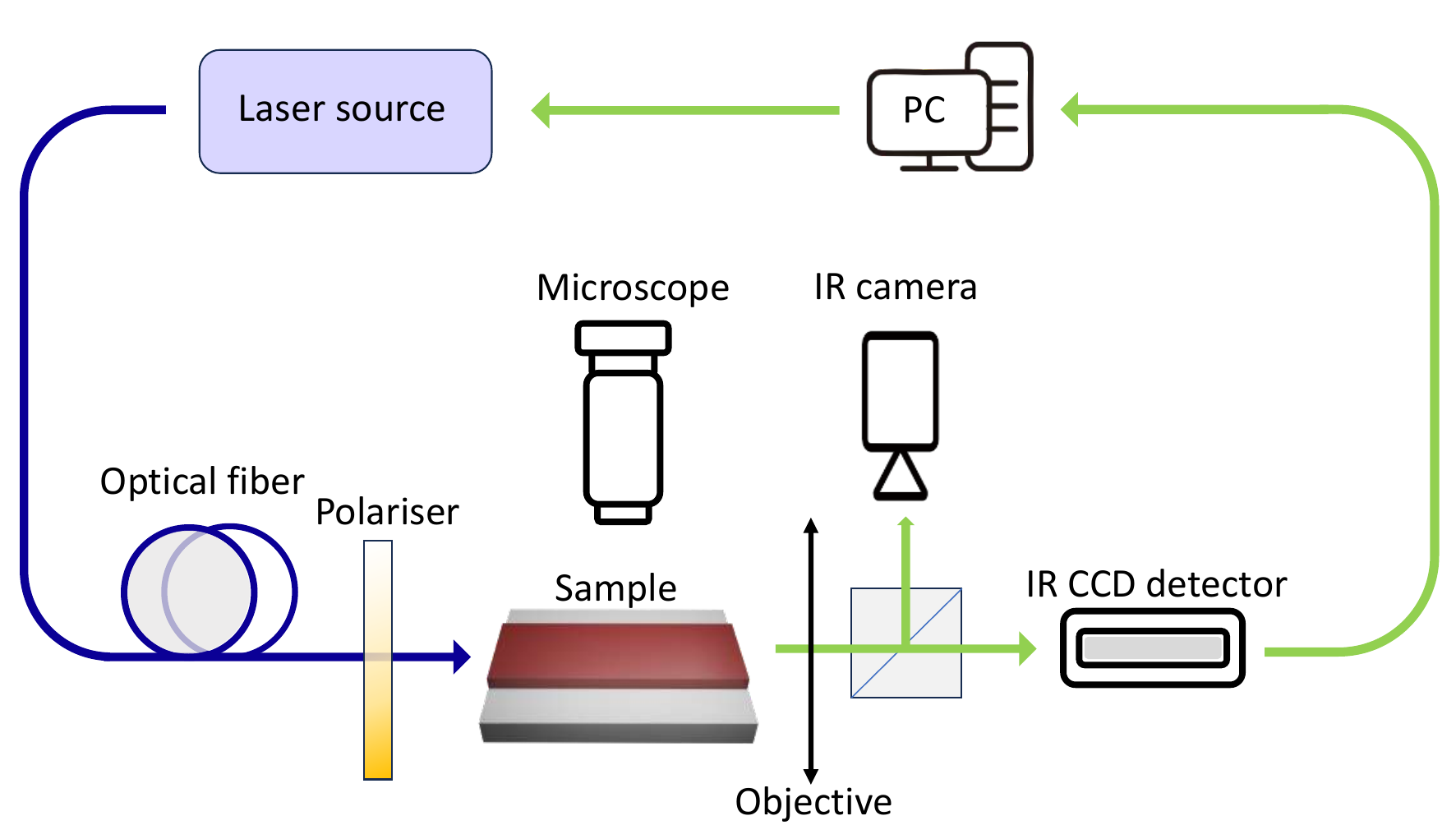}
        \caption{}
        \label{fig:setup1}
    \end{subfigure}
\hfill    ~ % ce symbole ajoute un espacement horizontal entre les premires deux images

    \begin{subfigure}[c]{0.95\textwidth}
        \centering \includegraphics[width=\textwidth]{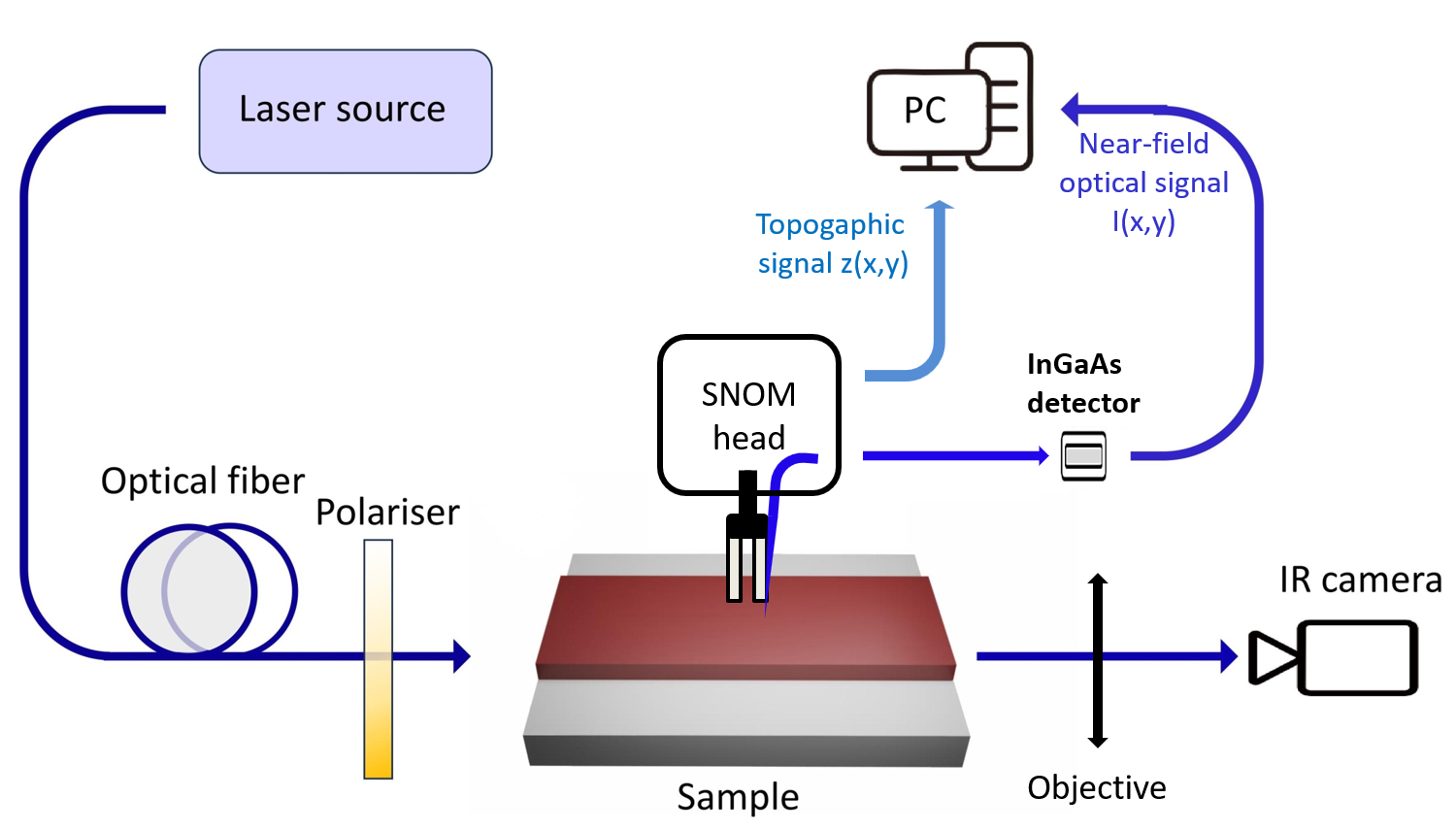}
        \caption{}
        \label{fig:setup2}
    \end{subfigure}
       \caption{Schematic diagrams of the two experimental setups. (a) End-fire optical bench setup for the structure with the fan-shaped set of output waveguides and  the 1D CCD camera. The sample is set on an alignment flexure stage for optical fiber coupling. It is aligned through the microscope and the IR camera. The polariser sets the incident electric field parallel to the plane of \Si layer. (b)  Setup for SNOM characterisation: it includes a tunable laser source, a lensed polarization-maintaining optical fibre, a control polariser, and the SNOM operating in collection mode with a metallized tapered optical fibre probe. The near-field optical signal is detected by an \ce{InGaAs} photodetector, while the sample topography is measured through the shear-force feedback system. A microscope objective and an IR camera are used to monitor light coupling and propagation in the the \Si layer.}
  \end{figure}

This measurement process has been previously described in\;\cite{apl124_yue}. 
The experimental setup with the CCD camera is shown in figure\;\ref{fig:setup1}. 
An end coupling system is used for the characterisation with output waveguides. Two tunable lasers of the TUNICS T100S series are employed as the input source, which covers a spectral range from 1380 to 1640\nm. A linearly polarised input wave is coupled into the sample using a polarisation maintaining lensed fiber. After passing through the HMFE, the output wave is collected using a 32$\times$ objective with 0.6 numerical aperture and is split into two. One part of the wave is collected by a large-area IR camera for simple alignement, and the other part is detected by a high-sensitivity 1D CCD array IR camera, which provides the spatial distribution of the output wave.

% ---------------------------------------------------------
% --- Characterisation using Scanning Near-Field Optical Microscopy
% ---------------------------------------------------------
\subsection{Characterisation using Scanning Near-Field Optical Microscopy}\label{sec:snom_method}
% ---------------------------------------------------------
% --- Version INL
% ---------------------------------------------------------
The optical field distribution generated by the Maxwell fish-eye flat lens was characterized using aperture scanning near-field optical microscopy (a-SNOM) operated in collection mode. The device was excited by a tunable laser source (Tunics Plus), allowing the wavelength to be varied from 1490 nm to 1590 nm. The polarization state of the injected light was controlled using a polarizer, enabling the selection of the transverse-electric (TE) polarization. Light was coupled into the \Si layer through the sample edge, with precise alignment ensured using a polarization-maintaining lensed fiber. After passing through the HMFE, the output beam is collected with a 20$\times$ objective (NA = 0.4) and imaged onto an infrared camera to optimize the coupling conditions.

A metallized tapered fiber probe was used to sample the evanescent tail of the guided field propagating within the \Si layer. Owing to the collection geometry and the probe characteristics, the recorded near-field signal is mainly associated with the electric-field components parallel to the sample surface (TE component), whereas the sensitivity to the normal component (Ez) is significantly reduced. A non-optical shear-force feedback loop maintained the probe at a distance of about 10\nm from the sample surface throughout the measurements. By raster-scanning the lens area, simultaneous maps of the near-field intensity and surface topography were acquired, providing direct access to the focusing behavior and spatial confinement of the guided field. The probe aperture was approximately 100\nm in diameter, ensuring subwavelength spatial resolution.

\section{Acknowledgements}

X.\;ZHENG warmly acknowledges the support of the China Scholarship Council (CSC). The authors thank the french RENATER network. 

\section{Author contributions statement}

X.Z., Y.Q. and E.A. proposed the concept. X.Z. and Y.Q designed the device and carried out the simulations. J.-R.C., X.Z. fabricated the samples. X.Z., A.L., A.B. and S.C. performed the experimental characterisation. X.Z and E.A. wrote the manuscript. All authors discussed the results and reviewed the manuscript. E.A. supervised the study. 

\section{Additional information}

%To include, in this order: \textbf{Accession codes} (where applicable); 
\textbf{Competing interests} (mandatory statement). The authors declare none. 

The corresponding author is responsible for submitting a \href{http://www.nature.com/srep/policies/index.html#competing}{competing interests statement} on behalf of all authors of the paper. This statement must be included in the submitted article file.

% --------------------------------------------------------
% --- Bibliographie
% --------------------------------------------------------

% \input{hmfe_xin_osa_submission.bbl}
%\bibliography{hmfe_xin_osa_submission.bib}

\begin{thebibliography}{10}
\newcommand{\enquote}[1]{``#1''}

\bibitem{xiang2021perspective}
C.~Xiang, S.~M. Bowers, A.~Bjorlin, R.~Blum, and J.~E. Bowers,
  \enquote{Perspective on the future of silicon photonics and electronics,}
  {\protect\JournalTitle{Applied Physics Letters}} \textbf{118} (2021).

\bibitem{engheta2007circuits}
N.~Engheta, \enquote{Circuits with light at nanoscales: optical nanocircuits
  inspired by metamaterials,} {\protect\JournalTitle{science}} \textbf{317},
  1698--1702 (2007).

\bibitem{thomson2016roadmap}
D.~Thomson, A.~Zilkie, J.~E. Bowers, T.~Komljenovic, G.~T. Reed, L.~Vivien,
  D.~Marris-Morini, E.~Cassan, L.~Virot, J.-M. F{\'e}d{\'e}li \emph{et~al.},
  \enquote{Roadmap on silicon photonics,} {\protect\JournalTitle{Journal of
  Optics}} \textbf{18}, 073003 (2016).

\bibitem{luque2019ultracompact}
J.~M. Luque-Gonz{\'a}lez, R.~Halir, J.~G. Wang{\"u}emert-P{\'e}rez,
  J.~de~Oliva-Rubio, J.~H. Schmid, P.~Cheben, {\'I}.~Molina-Fern{\'a}ndez, and
  A.~Ortega-Mo{\~n}ux, \enquote{An ultracompact {GRIN}-lens-based spot size
  converter using subwavelength grating metamaterials,}
  {\protect\JournalTitle{Laser \& Photonics Reviews}} \textbf{13}, 1900172
  (2019).

\bibitem{zhang2020ultra}
Y.~Zhang, Y.~He, H.~Wang, L.~Sun, and Y.~Su, \enquote{Ultra-broadband mode size
  converter using on-chip metamaterial-based luneburg lens,}
  {\protect\JournalTitle{ACS Photonics}} \textbf{8}, 202--208 (2020).

\bibitem{gomez2012gradient}
C.~Gomez-Reino, M.~V. Perez, and C.~Bao, \emph{Gradient-index optics:
  fundamentals and applications} (Springer Science \& Business Media, 2012).

\bibitem{bociort1994imaging}
F.~Bociort, \emph{Imaging properties of gradient-index lenses} (K{\"o}ster
  Berlin, 1994).

\bibitem{book_bornwolf}
M.~Born and E.~Wolf, \emph{Principles of Optics} (Cambridge University Press,
  1999), 7th ed.

\bibitem{maxwell1854solutions}
J.~C. Maxwell, \enquote{Solutions of problems,} {\protect\JournalTitle{Camb.
  Dublin Math. J.}} \textbf{9}, 9 -- 11 (1854).

\bibitem{njp13_tyc}
T.~Tyc, L.~Herzánová, M.~Šarbort, and K.~Bering, \enquote{Absolute
  instruments and perfect imaging in geometrical optics,}
  \href{http://dx.doi.org/10.1088/1367-2630/13/11/115004}{{\protect\JournalTitle{New
  Journal of Physics}}} \textbf{13}, 115004 (2011).

\bibitem{n7375_tyc_zhang}
T.~Tyc and X.~Zhang, \enquote{Perfect lenses in focus,}
  {\protect\JournalTitle{Nature}} \textbf{480}, 42--43 (2011).

\bibitem{fan2017integrated}
Y.~Fan, X.~Le~Roux, A.~Korovin, A.~Lupu, and A.~de~Lustrac, \enquote{Integrated
  {2D}-graded index plasmonic lens on a silicon waveguide for operation in the
  near infrared domain,} {\protect\JournalTitle{ACS nano}} \textbf{11},
  4599--4605 (2017).

\bibitem{ao19_moore}
D.~T. Moore, \enquote{Gradient-index optics: a review,}
  \href{http://dx.doi.org/10.1364/AO.19.001035}{{\protect\JournalTitle{Appl.
  Opt.}}} \textbf{19}, 1035--1038 (1980).

\bibitem{lei2017generalized}
Q.~Lei, R.~Foster, P.~S. Grant, and C.~Grovenor, \enquote{Generalized {Maxwell}
  fish-eye lens as a beam splitter: A case study in realizing all-dielectric
  devices from transformation electromagnetics,} {\protect\JournalTitle{IEEE
  Transactions on Microwave Theory and Techniques}} \textbf{65}, 4823--4835
  (2017).

\bibitem{headland2020half}
D.~Headland, M.~Fujita, and T.~Nagatsuma, \enquote{Half-{Maxwell} fisheye lens
  with photonic crystal waveguide for the integration of terahertz optics,}
  {\protect\JournalTitle{Optics express}} \textbf{28}, 2366--2380 (2020).

\bibitem{headland2021dielectric}
D.~Headland, A.~K. Klein, M.~Fujita, and T.~Nagatsuma, \enquote{Dielectric
  slot-coupled half-{Maxwell} fisheye lens as octave-bandwidth beam expander
  for terahertz-range applications,} {\protect\JournalTitle{APL Photonics}}
  \textbf{6} (2021).

\bibitem{bitton2018two}
O.~Bitton, R.~Bruch, and U.~Leonhardt, \enquote{Two-dimensional {Maxwell}
  fisheye for integrated optics,} {\protect\JournalTitle{Physical Review
  Applied}} \textbf{10}, 044059 (2018).

\bibitem{nanoph11_shen}
J.~Shen, Y.~Zhang, Y.~Dong, Z.~Xu, J.~Xu, X.~Quan, X.~Zou, and Y.~Su,
  \enquote{Ultra-broadband on-chip beam focusing enabled by grin metalens on
  silicon-on-insulator platform,}
  \href{http://dx.doi.org/https://doi.org/10.1515/nanoph-2022-0242}{{\protect\JournalTitle{Nanophotonics}}}
  \textbf{11}, 3603--3612 (2022).

\bibitem{centeno2005graded}
E.~Centeno and D.~Cassagne, \enquote{Graded photonic crystals,}
  {\protect\JournalTitle{Optics letters}} \textbf{30}, 2278--2280 (2005).

\bibitem{oc285_gaufillet}
F.~Gaufillet and {\'E}.~Akmansoy, \enquote{Graded photonic crystals for graded
  index lens,} {\protect\JournalTitle{Optics Communications}} \textbf{285},
  2638--2641 (2012).

\bibitem{om47_gaufillet}
F.~Gaufillet and {\'E}.~Akmansoy, \enquote{Design of flat graded index lenses
  using dielectric graded photonic crystals,} {\protect\JournalTitle{Optical
  Materials}} \textbf{47}, 555--560 (2015).

\bibitem{staude2017metamaterial}
I.~Staude and J.~Schilling, \enquote{Metamaterial-inspired silicon
  nanophotonics,} {\protect\JournalTitle{Nature Photonics}} \textbf{11},
  274--284 (2017).

\bibitem{cheben2018subwavelength}
P.~Cheben, R.~Halir, J.~H. Schmid, H.~A. Atwater, and D.~R. Smith,
  \enquote{Subwavelength integrated photonics,} {\protect\JournalTitle{Nature}}
  \textbf{560}, 565--572 (2018).

\bibitem{pj10_gaufillet}
F.~Gaufillet and {\'E}.~Akmansoy, \enquote{{Maxwell} fish-eye and
  half-{Maxwell} fish-eye based on graded photonic crystals,}
  {\protect\JournalTitle{IEEE Photonics Journal}} \textbf{10}, 1--10 (2018).

\bibitem{Lumerical}
Lumerical, \url{https://www.lumerical.com/products/fdtd}.

\bibitem{pj8_gaufillet}
F.~Gaufillet and {\'E}.~Akmansoy, \enquote{Graded photonic crystals for
  luneburg lens,} {\protect\JournalTitle{IEEE Photonics Journal}} \textbf{8},
  1--11 (2016).

\bibitem{yue2022dual}
Q.~Yue and {\'E}.~Akmansoy, \enquote{Dual-band flat lens with negative index
  for silicon photonics,} {\protect\JournalTitle{Applied Physics A}}
  \textbf{128}, 1--6 (2022).

\bibitem{jlt_xin}
X.~Zheng and {\'E}.~Akmansoy, \enquote{Effective index approximation method for
  graded photonic crystals slabs: Precise design of a {Mikaelian} lens,}
  \href{http://dx.doi.org/10.1109/JLT.2024.3426352}{{\protect\JournalTitle{Journal
  of Lightwave Technology}}} pp. 1--9 (2024).

\bibitem{acsam12_fan}
Y.~Fan, B.~Cluzel, M.~Petit, X.~Le~Roux, A.~Lupu, and A.~de~Lustrac,
  \enquote{{2D} waveguided bessel beam generated using integrated
  metasurface-based plasmonic axicon,}
  \href{http://dx.doi.org/10.1021/acsami.0c03420}{{\protect\JournalTitle{ACS
  Applied Materials \& Interfaces}}} \textbf{12}, 21114--21119 (2020). PMID:
  32310629.

\bibitem{apl124_yue}
Q.~Yue, X.~Leroux, B.~Cluzel, M.~Petit, A.~Lupu, and {\'E}.~Akmansoy,
  \enquote{{Graded flat lens with negative index for silicon photonics},}
  \href{http://dx.doi.org/10.1063/5.0195652}{{\protect\JournalTitle{Applied
  Physics Letters}}} \textbf{124}, 241102 (2024).

\bibitem{oe12_lupu}
A.~Lupu, E.~Cassan, S.~Laval, L.~E. Melhaoui, P.~Lyan, and J.~M. Fedeli,
  \enquote{Experimental evidence for superprism phenomena in {SOI} photonic
  crystals,}
  \href{http://dx.doi.org/10.1364/OPEX.12.005690}{{\protect\JournalTitle{Opt.
  Express}}} \textbf{12}, 5690--5696 (2004).

\bibitem{lee2018anderson}
M.~Lee, J.~Lee, S.~Kim, S.~Callard, C.~Seassal, and H.~Jeon, \enquote{Anderson
  localizations and photonic band-tail states observed in compositionally
  disordered platform,} {\protect\JournalTitle{Science Advances}} \textbf{4},
  e1602796 (2018).

\bibitem{vo2012near}
T.-P. Vo, M.~Mivelle, S.~Callard, A.~Rahmani, F.~Baida, D.~Charraut,
  A.~Belarouci, D.~Nedeljkovic, C.~Seassal, G.~Burr \emph{et~al.},
  \enquote{Near-field probing of slow {Bloch} modes on photonic crystals with a
  nanoantenna,} {\protect\JournalTitle{Optics express}} \textbf{20}, 4124--4135
  (2012).

\bibitem{fabre2008optical}
N.~Fabre, L.~Lalouat, B.~Cluzel, X.~M{\'e}lique, D.~Lippens, F.~de~Fornel, and
  O.~Vanb{\'e}sien, \enquote{Optical near-field microscopy of light focusing
  through a photonic crystal flat lens,} {\protect\JournalTitle{Physical review
  letters}} \textbf{101}, 073901 (2008).

\bibitem{aplp10_briche}
R.~Briche, A.~Benamrouche, P.~Cremillieu, P.~Regreny, J.-L. Leclercq,
  X.~Letartre, A.~Danescu, and S.~Callard, \enquote{Tubular optical
  microcavities based on rolled-up photonic crystals,}
  {\protect\JournalTitle{APL Photonics}} \textbf{5}, 106106 (2020).

\bibitem{oe33_kmiche}
M.~Kemiche, J.~Lhuillier, A.~Benamrouche, R.~Mazurczyk, P.~Regreny, T.~Wood,
  P.~Demongodin, S.~Callard, and C.~Monat, \enquote{Regular combs of modes in
  compact cavities based on slow-light dispersion engineering in active {III-V}
  planar photonic crystals,}
  \href{http://dx.doi.org/10.1364/OE.553385}{{\protect\JournalTitle{Opt.
  Express}}} \textbf{33}, 26100--26115 (2025).

\end{thebibliography}

\end{document}